\documentclass[twoside,twocolumn,english,floatfix, superscriptaddress, aps]{revtex4-2}
\usepackage[T1]{fontenc}
\usepackage[latin9]{inputenc}
\usepackage{amsbsy}
\usepackage{graphicx}

\makeatletter

\providecommand{\tabularnewline}{\\}

\date{\today}
\usepackage{amsmath}

\makeatother

\usepackage{babel}
\begin{document}
\author{Yingchao Lu}
\email{yingchao.lu@rice.edu}
\affiliation{Department of Physics and Astronomy, Rice University, Houston, Texas 77005, USA}
\affiliation{Theoretical Division, Los Alamos National Laboratory, Los Alamos, New Mexico, 87545, USA}
\author{Fan Guo}
\author{Patrick Kilian}
\author{Hui Li}
\author{Chengkun Huang}
\affiliation{Theoretical Division, Los Alamos National Laboratory, Los Alamos, New Mexico, 87545, USA}
\author{Edison Liang}
\affiliation{Department of Physics and Astronomy, Rice University, Houston, Texas 77005, USA}
\title{Fermi-type particle acceleration from magnetic reconnection at the
termination shock of a relativistic striped wind}
\begin{abstract}
An oblique-rotating pulsar generates a relativistic striped wind in
a pulsar wind nebula (PWN). The termination shock of the PWN compresses
the Poynting-flux-dominated flow and drives magnetic reconnection.
By carrying out particle-in-cell (PIC) simulations of the termination
shock of the PWN, we study the shock structure as well as the energy
conversion processes and particle acceleration mechanisms. With the
recent advances in the numerical methods, we extend the simulations
to the ultra-relativistic regime with bulk Lorentz factor up to $\gamma_{0}=10^{6}$.
Magnetic reconnection at the termination shock is highly efficient
at converting magnetic energy to particle kinetic energy and accelerating
particles to high energies. We find that the resulting energy spectra
crucially depend on $\lambda/d_{e}$. When $\lambda/d_{e}$ is large
($\lambda\gtrsim40d_{e}$) , the downstream particle spectra form
a power-law distribution in the magnetically dominated relativistic
wind regime with upstream magnetization parameter $\sigma_{0}=10$.
By analyzing particle trajectories and statistical quantities relevant
to particle energization, we find that Fermi-type mechanism dominates
the particle acceleration and power-law formation. We find that the
results for particle acceleration are scalable as $\gamma_{0}$ and
$\sigma_{0}$ increase to large values. The maximum energy for electrons
and positrons can reach hundreds of TeV if the wind has a bulk Lorentz
factor $\gamma_{0}\approx10^{6}$ and magnetization parameter $\sigma_{0}=10$,
which can explain the recent observations of high-energy gamma-rays
from pulsar wind nebulae (PWNe).
\end{abstract}
\maketitle

\section{Introduction}

A rotating pulsar creates the surrounding pulsar wind nebula (PWN)
by steadily releasing an energetic wind into the interior of the expanding
shockwave of a core-collapse supernova explosion \citep{Gaensler2006}.
The wind is composed of magnetized plasma of relativistic electrons
and positrons.  The wind propagates radially and abruptly transits
at the termination shock where the ram pressure balances that of the
surrounding medium (supernova remnant or interstellar medium depending
on the age and motion of the PWN \citep{Gaensler2006,ReviewSi2017}).
The wind can transit from being Poynting-dominated to being particle-dominated
at the termination shock. Particles in the pulsar wind will be accelerated
at the termination shock, producing a broadband spectrum that can
be observed from the radio to X-ray bands. The spectral breaks between
the radio and the X-ray band have been found in the synchrotron spectra
of pulsar wind nebulae (PWNe). Some of the spectral breaks suggest
emission processes fed by non-thermal particle distributions at the
termination shock \citep{TSacc_Rees1974,TSacc_Kennel1984,TSacc_MHD_Kennel1984}.
How the electromagnetic energy is converted into particle energy and
how non-thermal particles are efficiently accelerated at the termination
shock is not fully understood.

Recently, high-energy gamma-ray emissions have been detected in both
young PWNe and mid-aged PWNe. The Crab nebula (about $10^{3}$ years
old, and the Crab pulsar has spin-down luminosity $\dot{E}\sim5\times10^{38}\mathrm{erg/s}$)
is a prototype of young and energetic PWNe. The origin of the observed
photons of energy $E_{\mathrm{ph}}>100\ \mathrm{TeV}$ from the Crab
is likely due to the acceleration of leptons in the vicinity of PeV
in the Crab nebula \citep{VHE_Amenomori2019,VHE_Abeysekara2019}.
The ultra-high-energy electrons and positrons can be produced by particle
acceleration in the nebula \citep{Positron_Abeysekara2017}. Mid-age
pulsars such as Geminga (more than $10^{5}$ years old, $\dot{E}\sim3\times10^{34}\mathrm{erg/s}$)
are beyond the synchrotron cooling time, but still accelerate electrons
to very high energies in the nebulae \citep{Positron_CR_Yueksel2009}.
In a recent survey \citep{VHE_Abeysekara2020}, nine Galactic sources
are found to emit above 56 TeV with data from High Altitude Water
Cherenkov (HAWC) Observatory, eight of which are within a degree of
the Galactic plane (the ninth source is the Crab Nebula). Those eight
inner Galactic plane sources are associated with high spin-down pulsars
($\dot{E}\gtrsim10^{36}\mathrm{erg/s}$) and remain extended in apparent
size above 56 TeV even though the gamma-ray radiating electrons cool
quickly. How pulsar winds efficiently accelerate electrons and positrons
to high energies is a major puzzle and holds the key of understanding
the near-Earth positron anomaly \citep{Positron_CR_Yueksel2009,Positron_Accardo2014,Positron_Hooper2017}
as well as gamma rays from the Galactic Center \citep{CR_Abdo2007,Positron_CR_Yueksel2009,CR_Linden2018,VHE_Abeysekara2020}.

In the case of an oblique-rotating pulsar, a radially propagating
relativistic flow is continuously launched. Near the equatorial plane,
toroidal magnetic fields of alternating polarity, separated by current
sheets, are embedded in the flow. Such a flow has been modeled as
a steady state striped wind, containing a series of drifting Harries
current sheets \citep{Striped_Coroniti1990,Striped_Kirk2003}.  Numerical
simulations including magnetohydrodynamics (MHD) \citep{TSacc_MHD_Kennel1984,MHD_Porth2014,MHD_Olmi2015,MHD_Porth2016},
particle-in-cell (PIC) \citep{Sironi2011} and test-particle simulations
\citep{Trajectory_Giacinti2018} have been used to model the termination
shock of PWNe. However, how magnetic energy is converted, and the
role of the termination shock \citep{Shock_Sironi2013,Shock_Summerlin2011,Shock_Sironi2009}
and relativistic magnetic reconnection \citep{Reconn_Guo2014,Reconn_Sironi2014,Reconn_Guo2015,Reconn_Guo2016,Reconn_Guo2019}
in accelerating particles is still unclear \citep{ReviewSi2017}.
Magnetic reconnection driven by the termination shock may dissipate
the magnetic energy and accelerate particles \citep{Striped_Petri2007,Sironi2011}.

Particle acceleration in relativistic magnetic reconnection has been
a recent topic of strong interests. In the case of a spontaneous reconnection,
controversy on the role of direct acceleration and Fermi acceleration
in producing the power-law particle energy distribution has been extensively
addressed \citep{Reconn_Sironi2014,Reconn_Guo2014,Reconn_Guo2015}.
\citet{Reconn_Sironi2014} have suggested that the power-law forms
as the particles interact with the X-points (diffusion regions with
weak magnetic field $|\boldsymbol{E}|>|\boldsymbol{B}|$) through
direct acceleration. In contrast, analyses by \citet{Reconn_Guo2014,Reconn_Guo2015,Reconn_Guo2019}
show that the power-law distributions are produced by Fermi-like processes
and continuous injection from the reconnection inflow. In the case
of the shock-driven reconnection at the termination shocks of highly
relativistic striped pulsar winds, \citet{Sironi2011} have proposed
that high-energy particles are mainly accelerated at the electric
fields at the X-points. However, the role of Fermi-like processes
have not been studied in the shock-driven reconnection systems, which
is a main focus of this paper.

In this paper, we employ 2D PIC simulations to model the relativistic
striped wind interacting with the termination shock near the equatorial
plane of obliquely rotating pulsars. We focus on studying the dynamics
in a local box near the termination shock of the wind. While it is
extremely difficult to model the macroscopic system due to the enormous
scale separation between the system size and the skin depth, PIC models
provide a reliable and self-consistent description of the shock structure,
magnetic reconnection and particle acceleration.  We find that the
magnetic reconnection driven by the precursor perturbation from the
shock converts the magnetic energy into particle energy and accelerates
particles forming a power-law energy spectrum. We examine a wide range
of bulk Lorentz factor $10^{2}\le\gamma_{0}\le10^{6}$ and magnetization
parameter $10<\sigma_{0}\equiv B_{0}^{2}/(4\pi\gamma_{0}m_{e}n_{c0}c^{2})<300$
(assuming uniform magnetic field strength $B_{0}$ and uniform electron+positron
density $n_{c0}$ in the upstream) and show the scaling of the particle
spectrum. Most of our simulations have large $\gamma_{0}$ well above
the range $3<\gamma_{0}<375$ used in previous simulations \citep{Sironi2011}.
This is made possible using the recent improvement \citep{NCI_Lu2020}
of the PIC method to overcome the numerical problems. The wide range
of $\gamma_{0}$ and $\sigma_{0}$ is expected for PWNe with various
pulsar spin-down luminosity and age. Our analysis of the particle
trajectories and particle energization terms shows that Fermi-type
mechanisms by magnetic reconnection \citep{Reconn_Guo2019,Lemoine2019}
dominate the particle acceleration and power-law formation.

The rest of this paper is organized as follows. In Section \ref{sec:Numerical-simulations-setup},
we discuss the numerical methods and the setup of our simulations.
In Section \ref{sec:Shock-formation}, we discuss the evolution and
structure of the shock-reconnection system for the standard run. In
Section \ref{sec:Particle-spectrum}, we study the particle spectrum
and its dependency on parameters $\gamma_{0}$, $\sigma_{0}$ and
$\lambda$. In Section \ref{sec:Particle-acceleration-mechanism},
we give some detailed analyses of the particle acceleration mechanism.
We discuss and conclude the paper in Section \ref{sec:Conclusions-and-discussions}.

\section{Numerical simulations}

\label{sec:Numerical-simulations-setup}

\begin{figure}[tph]
\includegraphics[scale=0.56]{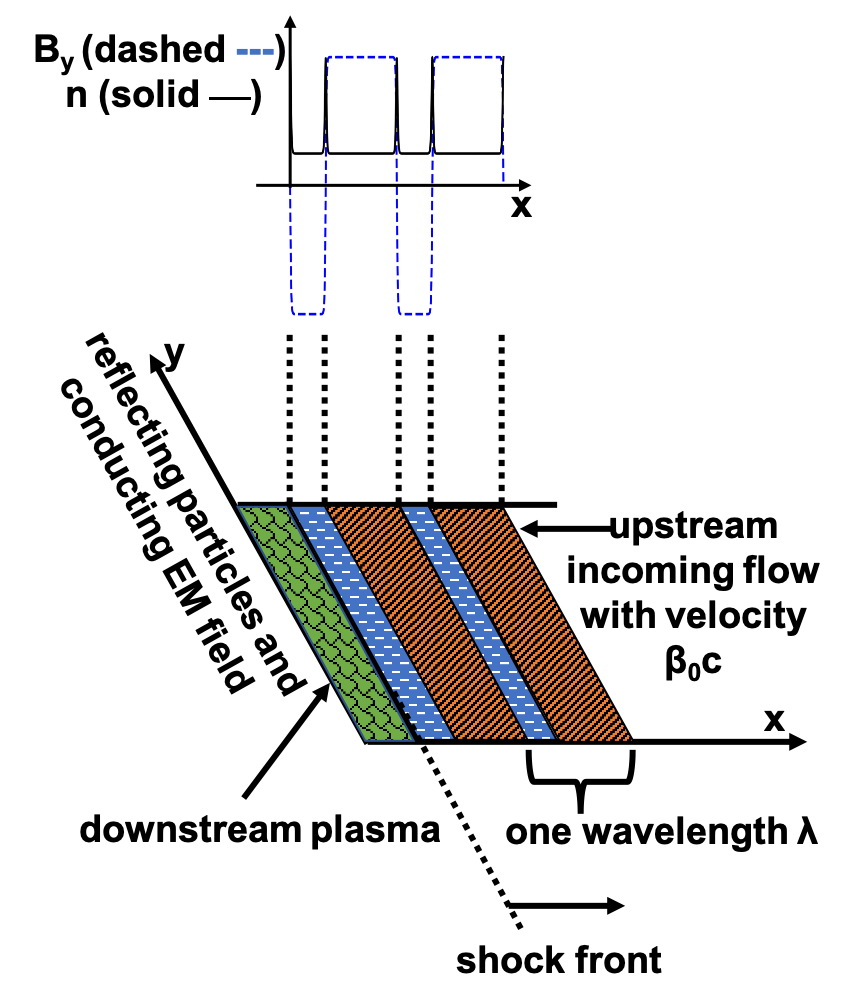}

\caption{2D PIC simulation setup for the termination shock of relativistic
striped wind. The upstream incoming flow drifts with bulk Lorentz
factor $\gamma_{0}$ and is composed of in-plane magnetic field $B_{y}$
in $+y$ or $-y$ direction, and particles (electrons and positrons).
In each current sheet, there is a hot dense component of particles
as shown in the density profile. The hot component balances the magnetic
pressure and ensures the steady electromagnetic profile. \label{fig:setup}}
\end{figure}

\begin{table}[tph]
\caption{Parameters for each run in this work. The parameters listed are the
transverse box size $L_{y}$, the wave length of the striped wind
$\lambda$, upstream bulk Lorentz factor $\gamma_{0}$, the thickness
of the current sheets $\Delta$ and the upstream magnetization $\sigma$.
The lengths are in the unit of skin depth $d_{e}$. \label{tab:Parameters}}

\hfill{}%
\begin{tabular}{|c|c|c|c|c|c|}
\hline 
Run & $L_{y}/d_{e}$ & $\lambda/d_{e}$ & $\gamma_{0}$ & $\Delta/d_{e}$ & $\sigma$\tabularnewline
\hline 
\hline 
S0 & 400 & 640 & $10^{4}$ & 1 & 10\tabularnewline
\hline 
A1 & 400 & 640 & $10^{2}$ & 1 & 10\tabularnewline
\hline 
A2 & 400 & 640 & $10^{3}$ & 1 & 10\tabularnewline
\hline 
A3 & 400 & 640 & $10^{5}$ & 1 & 10\tabularnewline
\hline 
A4 & 400 & 640 & $10^{6}$ & 1 & 10\tabularnewline
\hline 
B1 & $400\sqrt{3}$ & $640\sqrt{3}$ & $10^{4}$ & $\sqrt{3}$ & 30\tabularnewline
\hline 
B2 & $400\sqrt{10}$ & $640\sqrt{10}$ & $10^{4}$ & $\sqrt{10}$ & 100\tabularnewline
\hline 
B3 & $400\sqrt{30}$ & $640\sqrt{30}$ & $10^{4}$ & $\sqrt{30}$ & 300\tabularnewline
\hline 
C1 & 400 & 160 & $10^{4}$ & 1 & 10\tabularnewline
\hline 
C2 & 400 & 40 & $10^{4}$ & 1 & 10\tabularnewline
\hline 
C3 & 400 & 20 & $10^{4}$ & 1 & 10\tabularnewline
\hline 
\end{tabular}\hfill{}
\end{table}

\begin{figure*}[tph]
\hfill{}\includegraphics[scale=0.6]{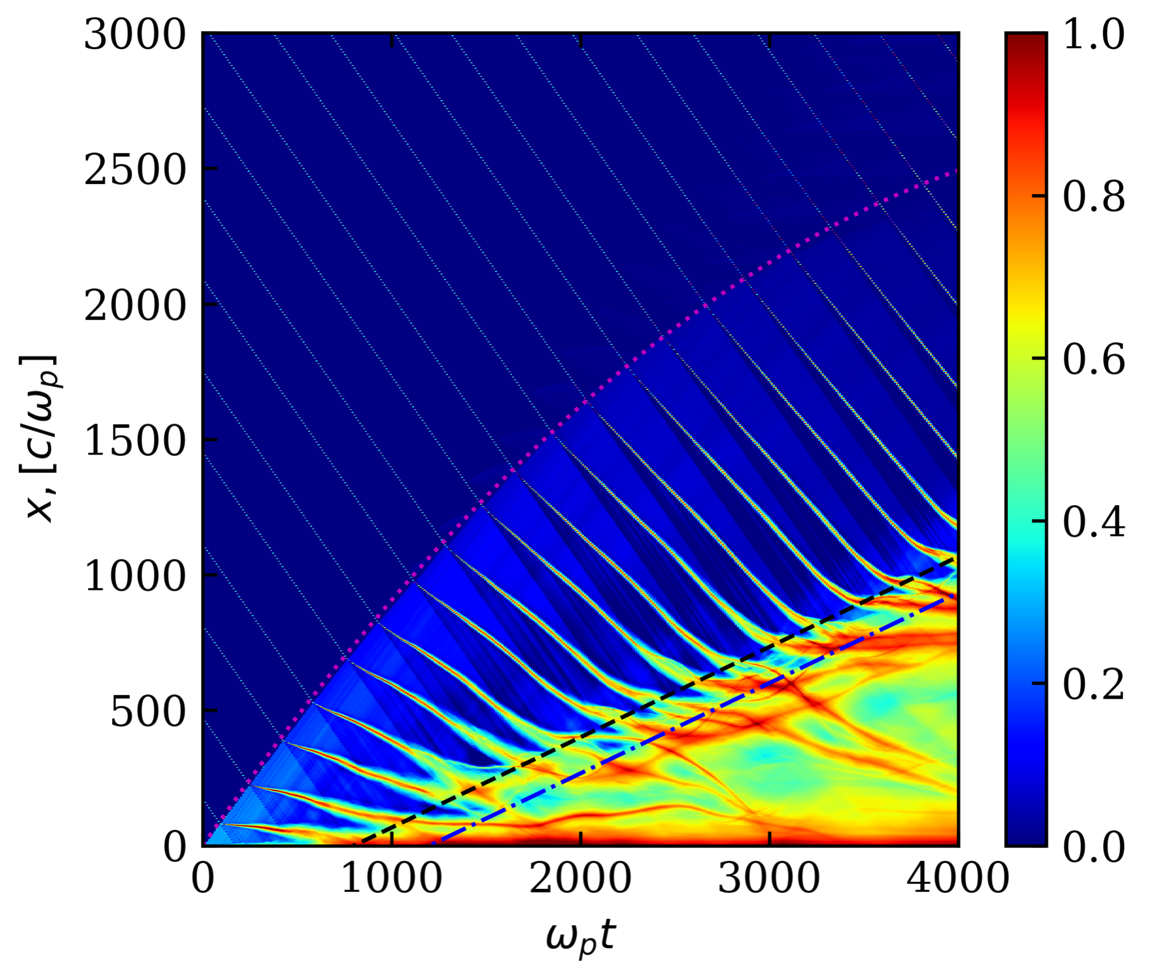}\hfill{}

\caption{The evolution of density profile averaged over $y$ direction, the
variable plotted is $\log_{10}(\langle n\rangle_{y}/n_{c0})$. The
magenta dotted line marks the location of the fast MHD shock (or the
precursor perturbation). The expression for the location of the fast
shock can be fitted as $x=ct[1-\omega_{p}t/103^{2}]$. The transition
region of the slow (main) shock is between the black dashed line and
the blue dot-dashed line. The expression for the transition region
is $(ct-1200c/\omega_{p})/3<x<(ct-800c/\omega_{p})/3$. \label{fig:density-1d}}
\end{figure*}

We use the 2D version of PIC code EPOCH \citep{EPOCH_Arber2015} to
study the structure and physical processes in the termination shock
of a relativistic striped wind. Overcoming numerical problems, especially
the numerical Cherenkov instability (NCI) \citep{NCI_Godfrey1974},
is critical for correctly modeling highly relativistic plasma flows.
To improve the numerical stability, we have heavily modified the code
to implement the WT (standing for \textbf{w}eighting with \textbf{t}ime-step
dependency) interpolation scheme \citep{NCI_Lu2020}, a piecewise
polynomial force interpolation scheme with time-step dependency. This
scheme eliminates the lowest order NCI growth rate and significantly
suppresses growth from the residue resonances of higher orders by
reducing time steps.

The spatial profile of the relativistic striped wind in our simulations
is shown in Figure \ref{fig:setup}. The steady electron-positron
flow propagates along $-x$ direction with a bulk Lorentz factor $\gamma_{0}$
before interacting with the reflected flow. The spatial profile of
the electromagnetic field in the simulation frame is
\begin{align}
B_{y}= & B_{0}\tanh\big\{\frac{1}{\delta}\big[\alpha+\cos\big(\frac{2\pi(x+\beta_{0}ct)}{\lambda}\big)\big]\big\}\\
E_{z}= & \beta_{0}B_{0}\tanh\big\{\frac{1}{\delta}\big[\alpha+\cos\big(\frac{2\pi(x+\beta_{0}ct)}{\lambda}\big)\big]\big\}
\end{align}
where $\beta_{0}$ is the velocity of the wind normalized by the speed
of light $c$, and $\lambda$ is the wavelength of the stripes in
the wind. The dimensionless parameters $\delta$ and $\alpha$ are
such that the half thickness of the current sheet is $\Delta=\lambda\delta/(2\pi)$,
and $B_{y}$ averaged over one wavelength is $\langle B_{y}\rangle_{\lambda}=B_{0}[1-2(\arccos\alpha)/\pi]$.
The background cold plasma in the wind is uniform, with constant density
$n_{e^{-},e^{+}}^{\mathrm{cold}}=n_{c0}/2$ and constant temperature
$kT_{e^{-},e^{+}}^{\mathrm{cold}}=0.04mc^{2}$ for both electrons
and positrons. The time in our simulations is normalized by $1/\omega_{p}$
where $\omega_{p}=\sqrt{4\pi n_{c0}e^{2}/(\gamma_{0}m_{e})}$ is the
plasma frequency, and the spatial coordinates in our simulation are
normalized by $d_{e}=c/\omega_{p}$. A hot electron-positron plasma
inside the Harris current sheets balances the magnetic pressure and
maintains the steady profile of electromagnetic field. The density
of the hot electron-positron plasma in the current sheet in the simulation
frame is

\begin{equation}
n_{e^{-},e^{+}}^{\mathrm{hot}}=\frac{n_{h0}}{2\cosh^{2}\big\{\frac{1}{\delta}\big[\alpha+\cos\big(\frac{2\pi(x+\beta_{0}ct)}{\lambda}\big)\big]\big\}}
\end{equation}
where $n_{h0}/n_{c0}=\eta$ is the over-density factor relative to
the cold particles outside the layer, and is set to be $\eta=3$ \citep{Striped_Kirk2003,Sironi2011,Reconn_Sironi2014}.
 The drift velocity in $z$ direction of the hot particles is setup
to ensure the steady profile of electromagnetic field, i.e. in the
rest frame of the wind $\nabla\times\boldsymbol{B}=(4\pi/c)\boldsymbol{J}$
is satisfied everywhere so that the electric field stays zero. The
left boundary located at $x=0$ is reflecting for particles and conducting
for electromagnetic fields. The shock is self-consistently generated
by the interaction between the reflected flow and the incoming flow.
Our simulations are performed in the downstream frame, where the resulting
downstream plasma bulk flow is at rest when the shock is well developed
and separated from the boundary. The simulation is periodic in the
$y$ direction. The length of the simulation box in the $y$ direction
is $L_{y}=400d_{e}$ for the standard run S0, which is large enough
to hold the largest island in the simulation and we have tested that
a larger length in $y$ direction does not change our main conclusions.
In our standard run, we have $\alpha=0.1$ (i.e. $\langle B_{y}\rangle_{\lambda}=0.064B_{0}$),
$\Delta=d_{e}$, $\lambda=640d_{e}$, $\gamma_{0}=1/\sqrt{1-\beta_{0}^{2}}=10^{4}$
and $\sigma_{0}=B_{0}^{2}/(4\pi\gamma_{0}m_{e}n_{c0}c^{2})=10$, and
$d_{e}$ is resolved with 7.5 computational cells. We use 4th order
particle shape, which significantly reduce the numerical noise even
for a relatively small number of particle per cell. Each computational
cell is initialized with two electrons and two positrons in the cold
wind, and additional two electrons and two positrons in the current
sheets. The parameters for other runs with different $\gamma_{0}$
and $\sigma_{0}$ are listed in Table \ref{tab:Parameters}, while
the resolution and $\alpha$ remain same for all the runs. We have
also performed limited experiments with higher resolutions, obtaining
essentially the same results.

To ensure that the onset of reconnection is independent of numerical
effects, we have extensively tested the stability of current sheets
in a double-periodic simulation without the reflecting wall. The results
using WT scheme \citep{NCI_Lu2020} are summarized in Appendix \ref{sec:Test-problem-periodic}.
It shows that the case with time step $\Delta t=0.2\Delta t_{\mathrm{CFL}}$
can make sure that the onset of reconnection is much later than $\omega_{p}t=2000$,
which is roughly the longest time it takes for the shock to compress
a current sheet in our simulation. Therefore, we choose $\Delta t=0.2\Delta t_{\mathrm{CFL}}$
for all runs in this paper (see Table \ref{tab:Parameters}).

\section{Shock formation and evolution}

\label{sec:Shock-formation}

\begin{figure*}[tph]
\includegraphics[scale=0.53]{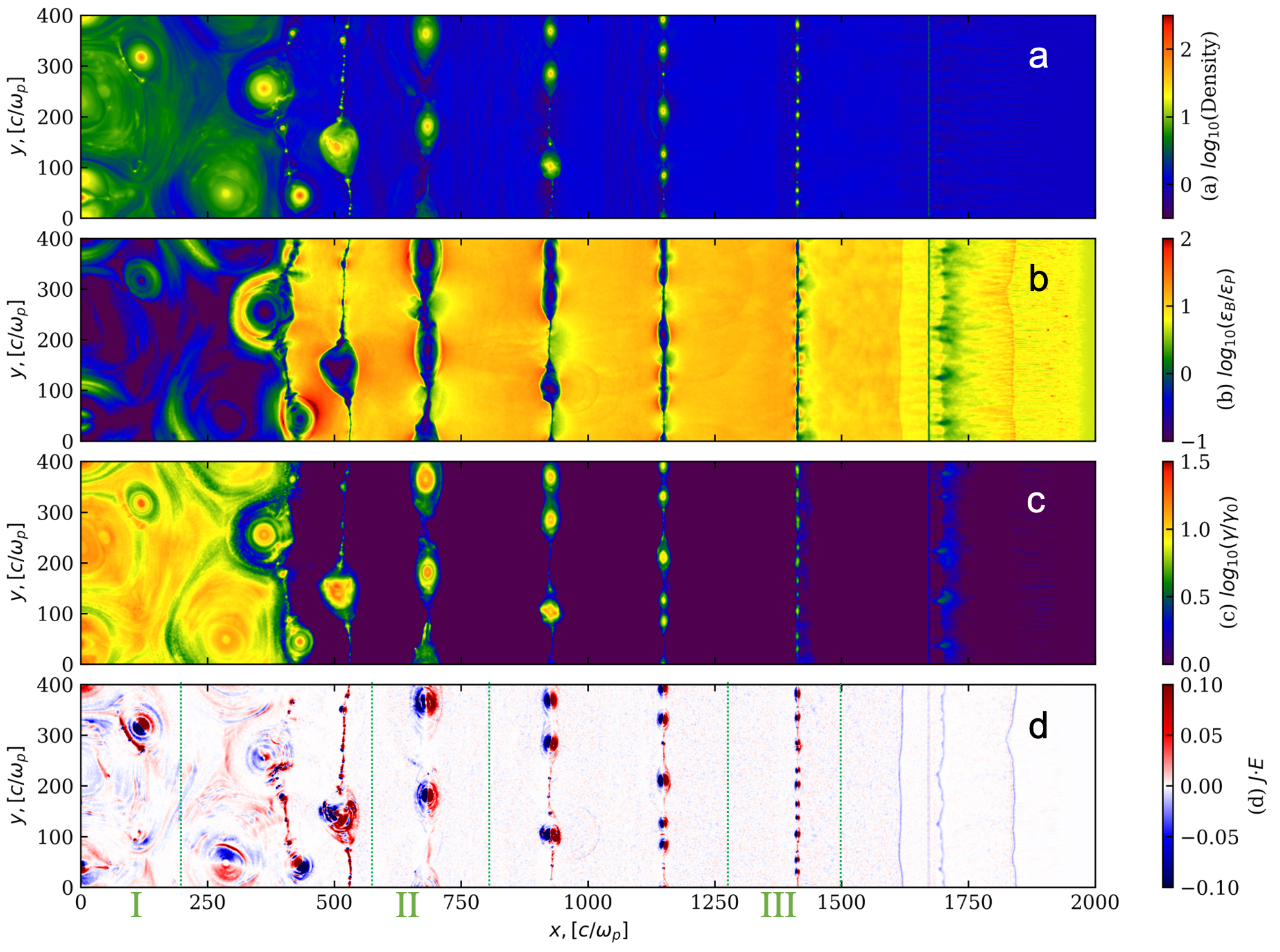}

\caption{2D spatial profile of different quantities for the standard run S0
with $\sigma_{0}=10$, $\gamma_{0}=10^{4}$, $\alpha=0.1$ at time
$\omega_{p}t=2000$. (a) logarithm of particle number density normalized
by $n_{c0}$, (b) logarithm of the ratio of magnetic energy density
to particle energy density, (c) logarithm of average $\gamma/\gamma_{0}$,
(d) energy conversion rate $\boldsymbol{J}\cdot\boldsymbol{E}$. The
green dotted vertical lines are boundaries of regions for more detailed
studies, region I for $0<x/d_{e}<197$, region II for $573<x/d_{e}<804$,
and region III for $1275<x/d_{e}<1497$. \label{fig:plot2D}}
\end{figure*}

In Figure \ref{fig:density-1d}, we show the evolution of density
profile averaged over the $y$ direction. In Figure \ref{fig:plot2D},
we show the 2D spatial distribution of density $\log_{10}(n/n_{c0})$,
magnetic energy density over particle energy density $\log_{10}(\varepsilon_{B}/\varepsilon_{P})$,
averaged particle kinetic energy $\gamma/\gamma_{0}$, and particle
energization rate $\boldsymbol{J}\cdot\boldsymbol{E}$ at $\omega_{p}t=2000$.
The shock-reconnection system forms and evolves self-consistently
as the reflected plasma interacts with the incoming flow. During this
initial process we observe a precursor perturbation, which is actually
a fast shock, propagates towards the upstream side of the simulation
box, compresses and decelerates the incoming currents sheets. This
drives the magnetic reconnection because the perturbed and compressed
current sheets are subject to rapid growth of the tearing instability.
Magnetic reconnection rapidly converts magnetic energy into particle
kinetic energies as current sheets evolve. This is further facilitated
by the compression at the main shock layer, characterized by a huge
density increase. The general shock and reconnection dynamics of our
standard run S0 are similar to the previous study \citep{Sironi2011},
where a fast shock (the precursor perturbation) and a slow hydrodynamic
shock (main shock) are identified.

The fast shock travels close to the speed of light initially and slows
down when interacting with the incoming flow afterwards, as tracked
by the magenta dotted trajectory in Figure \ref{fig:density-1d}.
The location of the precursor can be fitted as $x=ct[1-\omega_{p}t/103^{2}]$,
with the region with larger $x$ being unperturbed. The jump at the
fast shock is sharp in the beginning and becomes smoother after it
encounters more current sheets. The incoming current sheets are slowed
down and compressed as they pass through the fast shock. As shown
in Figure \ref{fig:plot2D}(b), the ratio of magnetic energy density
to particle energy density $\varepsilon_{B}/\varepsilon_{P}$ in the
regions between current sheets increases right after passing through
the fast shock, indicating strong compression. As a result of the
slowdown and weakening of the fast shock, a current sheet reaching
the fast shock later gets slowed down less than the one reaching the
fast shock earlier. More importantly, the fast shock compresses the
current sheets, triggering the onset of fast reconnection (also see
discussion below). There is a huge density jump at the main shock
around $x_{\mathrm{sh}}\approx354c/\omega_{p}$ at $\omega_{p}t=2000$.
The jump moves at a constant speed approximately $c/3$ (consistent
with the jump condition of ultra-relativistic shock) in $+\hat{\boldsymbol{x}}$
direction as shown in Figure \ref{fig:plot2D}(a). This density jump
can also be seen in Figure \ref{fig:density-1d} between the black
dashed line and the blue dot-dashed line, i.e. $(ct-1200c/\omega_{p})/3<x_{\mathrm{sh}}<(ct-800c/\omega_{p})/3$.

Inside the transition region between the precursor and the slow shock,
i.e. between the magenta dotted line and the black dashed line in
Figure \ref{fig:density-1d} or $354c/\omega_{p}\lesssim x\lesssim1623c/\omega_{p}$
in Figure \ref{fig:plot2D}, the adjacent regions of opposite magnetic
field polarity are pushed toward each other and reconnects. The current
sheets continuously break into a series of magnetic islands separated
by X-points. Strong energy conversion is associated with magnetic
reconnection, resulting in small $\varepsilon_{B}/\varepsilon_{P}$
in the islands, as shown in \ref{fig:plot2D}(c). The islands coalesce,
grow to larger sizes as they continuously move towards the downstream,
and further grow after passing the main shock front. In the downstream,
the average $\varepsilon_{B}/\varepsilon_{P}$ is around 0.3, much
less than its initial value $\varepsilon_{B}/\varepsilon_{P}=5$,
indicating that most of the magnetic energy is converted to particle
kinetic energy. As shown in Figure \ref{fig:plot2D}(c),  the average
particle Lorentz factor is $\langle\gamma\rangle\sim\gamma_{0}(1+\sigma_{0})=11\gamma_{0}$
in the downstream. The particles in the downstream can form thermal
and non-thermal distributions, which will be discussed in Section
\ref{sec:Particle-spectrum}. The energization rate $\boldsymbol{J}\cdot\boldsymbol{E}$,
as plotted in Figure \ref{fig:plot2D}(d), are concentrated in the
current sheets and islands. More details about the energization will
be discussed in Section \ref{sec:Particle-acceleration-mechanism}.

\section{Particle spectrum}

\label{sec:Particle-spectrum}

\begin{figure}[tph]
\includegraphics[scale=0.55]{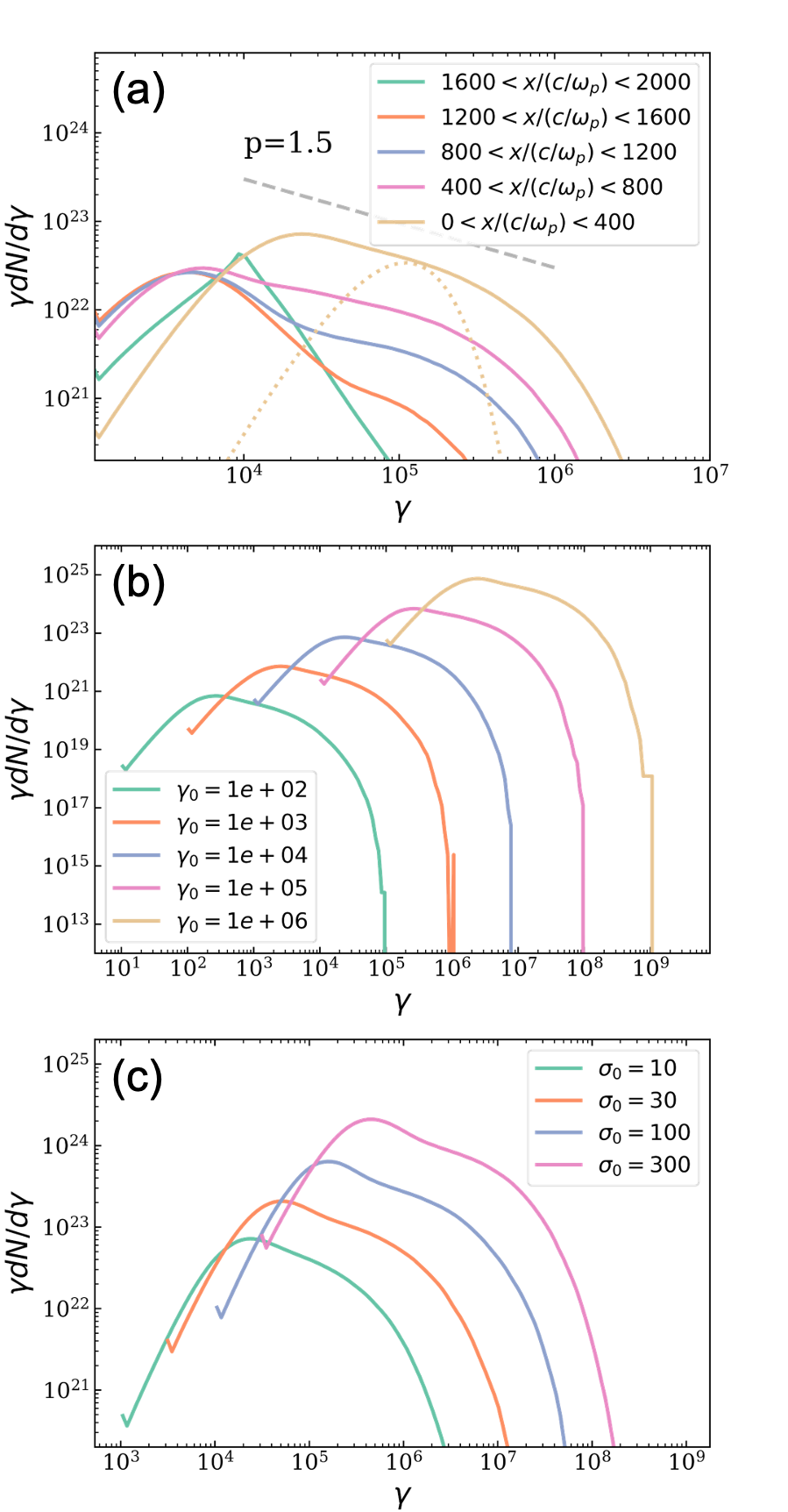}

\caption{Particle energy distribution $\gamma dN/d\gamma$ for all the electrons
and positrons in different regions and different runs (a) At $\omega_{p}t=2000$
for the standard run S0. The dashed line is the relativistic thermal
distribution for the downstream with peak at $\gamma_{0}(1+\sigma_{0})=1.1\times10^{5}$.
(b) For runs S0, A1, A2, A3, A4 (with different $\gamma_{0}$) in
downstream region at time $\omega_{p}t=20\sqrt{\gamma_{0}}$. (c)
For runs S0, B1, B2, B3 (with different $\sigma_{0}$) in downstream
region at time $\omega_{p}t=632\sqrt{\sigma_{0}}$. \label{fig:plot-spect1}}
\end{figure}

\begin{figure*}[tph]
\includegraphics[scale=0.61]{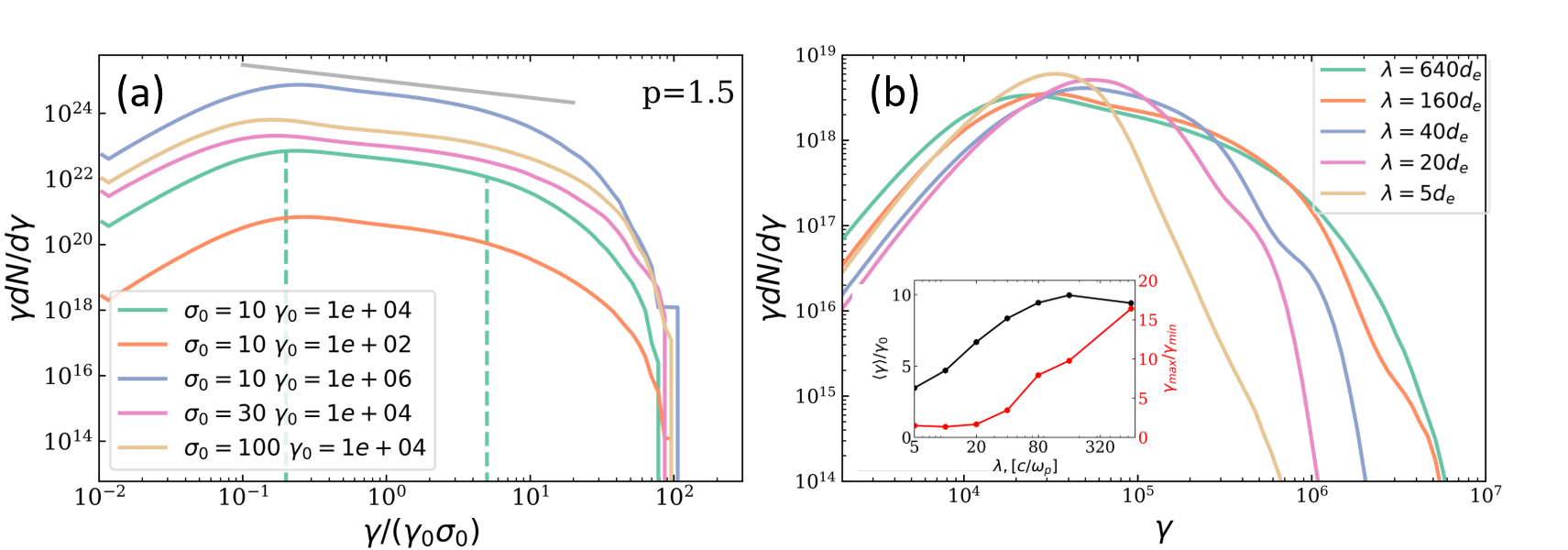}

\caption{Particle energy distribution $\gamma dN/d\gamma$ for electrons and
positrons in different runs (a) For runs S0, A2, A4, B1, B2 in downstream
region at time $\omega_{p}t=6.32\sqrt{\gamma_{0}\sigma_{0}}$. Note
that the horizontal axis is $\gamma/(\sigma_{0}\gamma_{0})$ and the
vertical dashed lines are at $\gamma/(\gamma_{0}\sigma_{0})=0.2$
and $\gamma/(\gamma_{0}\sigma_{0})=4$. (b) For runs S0, C1, C2, C3
(with different wavelength $\lambda$) in downstream region. In the
subpanel, the black line shows the average Lorentz factor of downstream
particles for different $\lambda$, the red line presents $\gamma_{\mathrm{max}}/\gamma_{\mathrm{min}}$
for different $\lambda$, where $\gamma_{\mathrm{min}}$ is defined
as the $\gamma$ where $\gamma dN/d\gamma$ peaks and $\gamma_{\mathrm{max}}$
is defined as the $\gamma$ where $\gamma^{2}dN/d\gamma$ peaks.\label{fig:plot-spect2}}
\end{figure*}

As the shock forms and interacts with the current sheets embedded
in the striped wind, the electrons and positrons are heated and accelerated.
In Figure \ref{fig:plot-spect1}(a), we show the energy distribution
function $f(\gamma)$ (related to the number distribution function
$n(\gamma)$ by $f(\gamma)\equiv n(\gamma)\gamma$) in different regions
for the standard run S0 at $\omega_{p}t=2000$. Before reaching the
fast shock, i.e. for $1600<x/(c/\omega_{p})<2000$, $f(\gamma)$ has
a peak at the bulk Lorentz factor $\gamma_{0}=10^{4}$. As the flow
slows down, i.e. for $400<x/(c/\omega_{p})<1600$, the peak of the
spectrum shifts to a lower energy. As the current sheets reconnect,
the tail of the spectrum becomes harder as the flow gets closer to
the main shock. According to the jump condition, the averaged kinetic
energy due to the energization in the downstream region is $\langle\gamma\rangle\approx\gamma_{0}(1+\sigma_{0})$.
For instance, in the standard run S0 ($\gamma_{0}=10^{4}$ and $\sigma_{0}=10$),
we have $\langle\gamma\rangle\approx1.1\times10^{5}$, which is consistent
with the energy density for $0<x/(c/\omega_{p})<400$ in Figure \ref{fig:plot-spect1}(a).
However, the actual downstream particle spectrum is distinctly different
from the thermal distribution. The particles with energies between
$\gamma=2\times10^{4}$ and $\gamma=4\times10^{5}$ follow a power-law
distribution $f(\gamma)\propto\gamma^{-p+1}$ with $p=1.5$. For comparison,
we also plot the distribution of a thermal plasma with $\langle\gamma\rangle=1.1\times10^{5}$
as the dashed line.

How does the resulting energy spectra depend on $\gamma_{0}$, $\sigma_{0}$
and $\lambda$? Earlier numerical simulations \citep{Sironi2011}
are limited to regimes with $\gamma_{0}\le375$. However, global models
\citep{Striped_Kirk2003,Flares_Kirk2017} have suggested a much larger
$\gamma_{0}$ and wider range of $\sigma_{0}$. Thanks to our advances
in numerical scheme \citep{NCI_Lu2020} that substantially reduces
the growth of numerical instabilities, we are able to explore a much
larger range of parameters. For the runs with different $\gamma_{0}$
and $\sigma_{0}$ in the range $10^{2}\le\gamma_{0}\le10^{6}$ as
shown in Figure \ref{fig:plot-spect1}(b) and in the range $10<\sigma_{0}<300$
as shown in \ref{fig:plot-spect1}(c), the downstream particle energy
distributions are similar to the one in the standard run. We confirmed
that the power-law energy distribution is scalable for different $\gamma_{0}$
and $\sigma_{0}$. The particles with energy $0.2<\gamma/(\gamma_{0}\sigma_{0})<4$
follow a power-law distribution with $p=1.5$. For different values
of magnetization parameter $\sigma_{0}$ and initial bulk flow Lorentz
factor $\gamma_{0}$, the spectra are similar if one plots the spectrum
against $\gamma/(\sigma_{0}\gamma_{0})$ and the variables $\lambda/(\sqrt{\sigma}c/\omega_{p})$
and $\alpha$ are kept the same for different runs, as shown in Figure
\ref{fig:plot-spect2}(a).  Despite the wide ranges of $\gamma_{0}$
and $\sigma_{0}$, the dimensionless particle-field equations and
the initial conditions are the same, as derived in Appendix \ref{sec:Scaling-equations-for}.
Thus the solutions, i.e. including the particle and field distributions,
between different $\gamma_{0}$ and $\sigma_{0}$ are scalable. The
particle spectrum depends on the value of $\lambda/(c/\omega_{p})$,
as shown in Figure \ref{fig:plot-spect2}(b). While the mean energy
per particle does not appreciably vary with $\lambda$ as shown by
the red line in the subpanel of Figure \ref{fig:plot-spect2}(b),
smaller $\lambda$ results in narrower spectrum and softer high-energy
tail. The ratio $\gamma_{\mathrm{max}}/\gamma_{\mathrm{min}}$ increases
as $\lambda$ increases and is significantly larger than unity for
$\lambda\gtrsim40d_{e}$. The power law spectrum forms for the runs
with $\lambda\gtrsim40d_{e}$. This is similar to the previous study
\citep{Sironi2011} and will be further explained in Section \ref{sec:Particle-acceleration-mechanism}.

The power-law distribution is stable as long as the shock is well
developed and separated from the boundary. The lower bound energy
$\varepsilon_{\mathrm{lb}}$ and the maximum energy $\varepsilon_{\mathrm{max}}$
scale with $\gamma_{0}$ and $\sigma_{0}$ as $\varepsilon_{\mathrm{lb}}=0.2\gamma_{0}\sigma_{0}$
and $\varepsilon_{\mathrm{max}}=4\gamma_{0}\sigma_{0}$. The maximum
energy for electrons and positrons can reach hundreds of TeV ($\gamma>2\times10^{8}$)
for the run with $\gamma_{0}\approx10^{6}$ and $\sigma_{0}=10$.
The power-law distribution of the particles suggests that an acceleration
mechanism is very efficient at energizing particles, which will be
analyzed and discussed in Section \ref{sec:Particle-acceleration-mechanism}.

\section{Particle acceleration mechanism}

\label{sec:Particle-acceleration-mechanism}

\begin{figure*}[tph]
\includegraphics[scale=0.95]{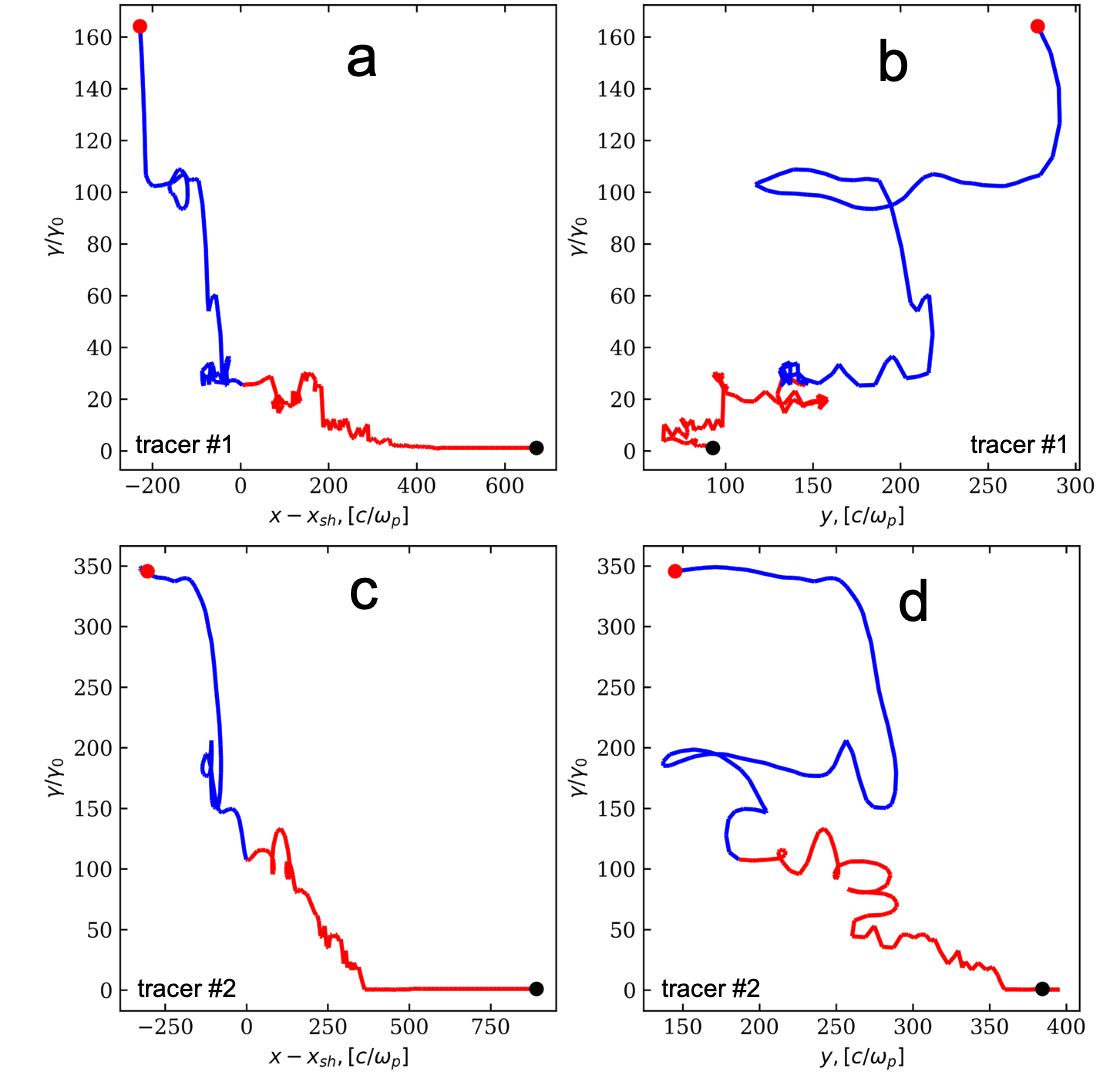}

\caption{Particle trajectories for two different particles. The left column
(a) and (c) is particle Lorentz factor (normalized by $\gamma_{0}$)
vs. $x-x_{\mathrm{sh}}$ coordinate where the location of the shock
jump is approximately at $x_{\mathrm{sh}}\approx(ct-1000c/\omega_{p})/3$,
and the right column (b) and (d) is particle Lorentz factor (normalized
by $\gamma_{0}$) vs. $y$ coordinate. The (a) and (b) are for tracer
particle \#1, and (c) and (d) are for tracer particle \#2. The red
line is the trajectory where the particle is in the upstream, and
the red line is the trajectory where the particle is in the downstream.
The black dots are for particles at $t=0$, and the red dots are for
particles at $\omega_{p}t=2000$. The main acceleration process is
Fermi-like bounces and particles gain significant amount of energy
during each bounce.\label{fig:plot-acc-1}}
\end{figure*}

\begin{figure*}[tph]
\includegraphics[scale=0.52]{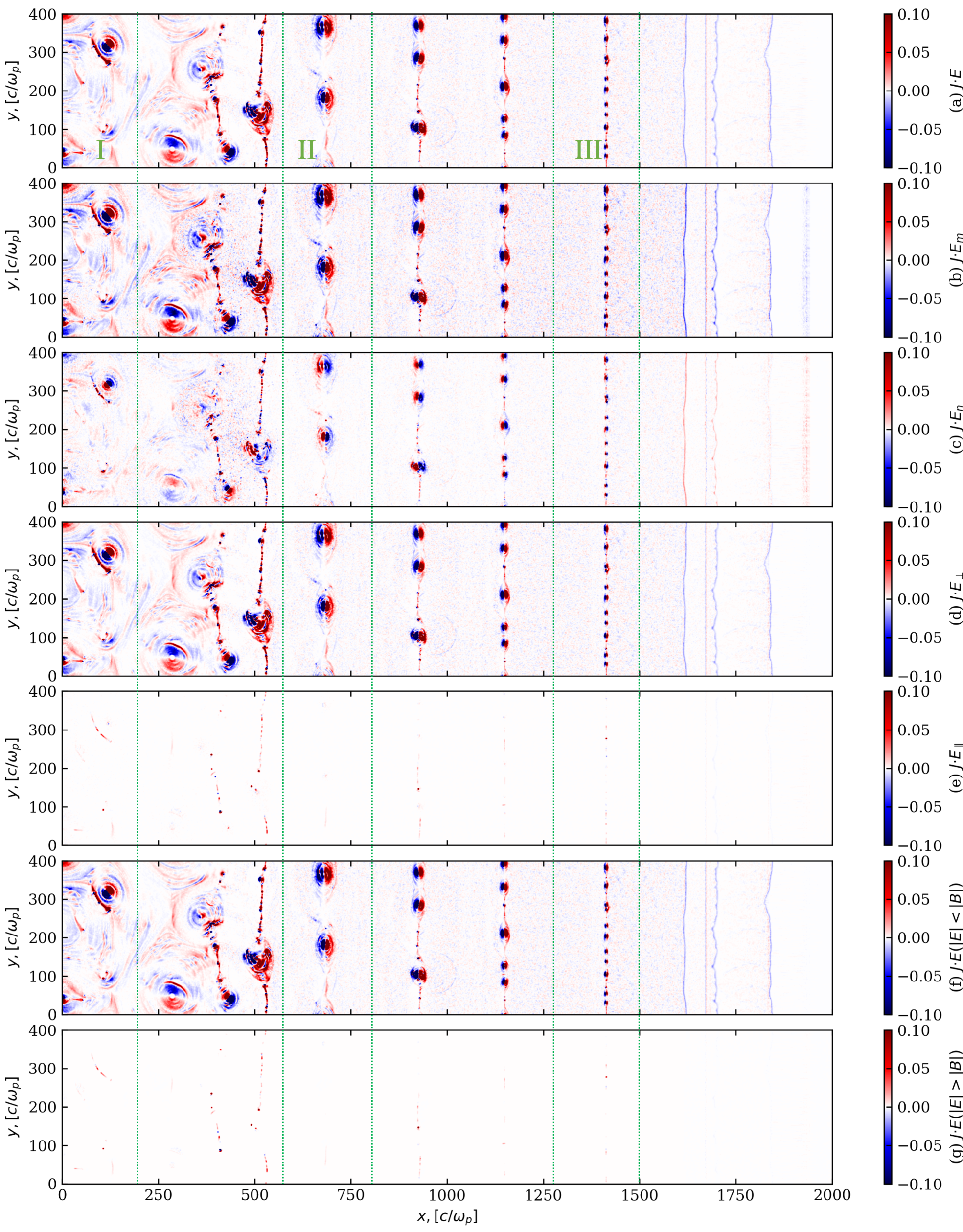}

\caption{2D spatial profile for $\boldsymbol{J}\cdot\boldsymbol{E}$ terms
in the standard run S0 with $\sigma_{0}=10$, $\gamma_{0}=10^{4}$,
$\alpha=0.1$ at time $\omega_{p}t=2000$. (a) total energization
rate $\boldsymbol{J}\cdot\boldsymbol{E}$, (b) $\boldsymbol{J}\cdot\boldsymbol{E}_{m}$
where $\boldsymbol{E}_{m}$ is the motional electric field, (c) $\boldsymbol{J}\cdot\boldsymbol{E}_{n}$
where $\boldsymbol{E}_{n}$ is the non-ideal electric field, (d) $\boldsymbol{J}\cdot\boldsymbol{E}_{\perp}$
where $\boldsymbol{E}_{\perp}$ is the generalized perpendicular electric
field, (e) $\boldsymbol{J}\cdot\boldsymbol{E}_{\parallel}$ where
$\boldsymbol{E}_{\parallel}$ is the generalized parallel electric
field, (f) $\boldsymbol{J}\cdot\boldsymbol{E}$ for $|\boldsymbol{E}|<|\boldsymbol{B}|$,
(g) $\boldsymbol{J}\cdot\boldsymbol{E}$ for $|\boldsymbol{E}|>|\boldsymbol{B}|$.
All the subfigures have the same color scale. $\boldsymbol{J}\cdot\boldsymbol{E}_{n}$
has smaller value but opposite polarity compared with $\boldsymbol{J}\cdot\boldsymbol{E}_{m}$.
$\boldsymbol{J}\cdot\boldsymbol{E}_{\perp}$ and $\boldsymbol{J}\cdot\boldsymbol{E}(|\boldsymbol{E}|<|\boldsymbol{B}|)$
are greater than $\boldsymbol{J}\cdot\boldsymbol{E}_{\parallel}$
and $\boldsymbol{J}\cdot\boldsymbol{E}(|\boldsymbol{E}|>|\boldsymbol{B}|)$.
\label{fig:JdotE}}
\end{figure*}

\begin{figure}[tph]
\includegraphics[scale=0.83]{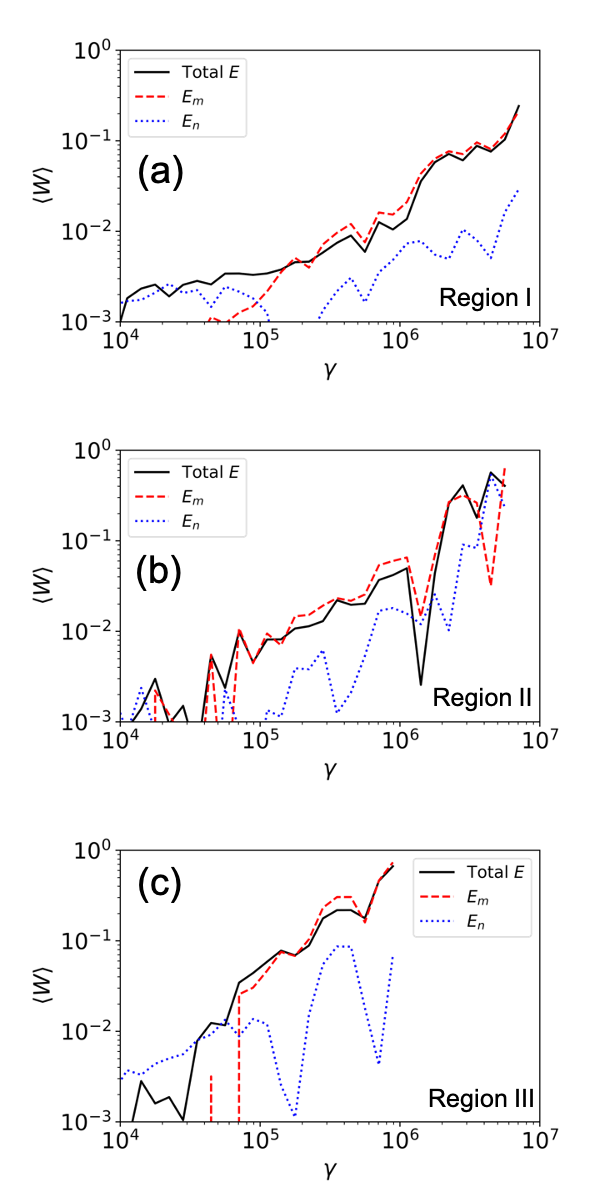}

\caption{(a) Acceleration rate by total electric field $\boldsymbol{E}$, motional
electric field $\boldsymbol{E}_{m}$ and non-ideal electric field
$\boldsymbol{E}_{n}$ for particles at different energy for $0<x<197c/\omega_{p}$,
i.e. region I in Figure \ref{fig:JdotE}, at $\omega_{p}t=2000$.
Note that the curve for $E-E_{m}$ is the absolute value. (b) same
as (a) but for $573c/\omega_{p}<x<804c/\omega_{p}$, i.e. region II
in Figure \ref{fig:JdotE}. (c) same as (a) but for $1275c/\omega_{p}<x<1497c/\omega_{p}$,
i.e. region III in Figure \ref{fig:JdotE}. $\boldsymbol{E}_{n}$
only contributes to positive acceleration in a small subset of energy
ranges especially in low energies compared to the contribution from
$\boldsymbol{E}_{m}$. \label{fig:plot-acc-a}}
\end{figure}

\begin{figure}[tph]
\includegraphics[scale=0.83]{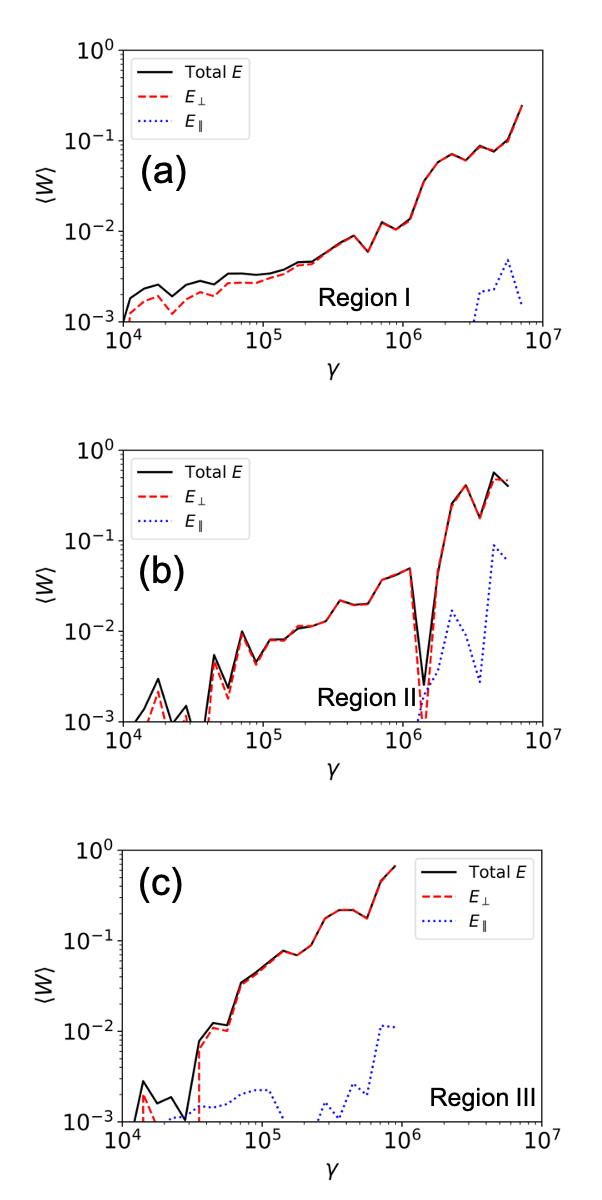}

\caption{Same as Figure \ref{fig:plot-acc-a} but for the generalized perpendicular
electric field $\boldsymbol{E}_{\perp}$ and parallel electric field
$\boldsymbol{E}_{\parallel}$. $\boldsymbol{E}_{\parallel}$ only
contributes to positive acceleration in a small subset of energy ranges
in high energies compared to the contribution from $\boldsymbol{E}_{\perp}$.
\label{fig:plot-acc-b}}
\end{figure}

In order to understand the particle acceleration mechanism at the
termination shock of the striped wind, we adopt several techniques
for analyzing the results of the standard run, including the particle
trajectory, decomposing the particle energization $\boldsymbol{J}\cdot\boldsymbol{E}$
term, and the acceleration rate binned by particle kinetic energy
for distinguishing different acceleration mechanisms. With these analyses,
we find that the major acceleration mechanism is a Fermi-type acceleration.
In the past, the diagnostics for understanding particle acceleration
often relies on a few hand-selected particles. Our analysis not only
includes particle trajectories but also statistically evaluate the
acceleration rate in a quantitative way, so we can evaluate competitive
processes without biases.

In Figure \ref{fig:plot-acc-1}, we show the trajectories for two
typical tracer particles in the simulation. Many of the trajectories
we get from the standard run are Fermi-like especially for those accelerated
to very high energies, i.e. particles bouncing back and forth and
gaining energy. The most important acceleration location is not at
the reconnection X-points, but by the relativistic flows generated
during magnetic reconnection.  As shown in Figure \ref{fig:plot2D}(d),
the energization rate $\boldsymbol{J}\cdot\boldsymbol{E}$ near the
X-points is relatively small compared to that in the reconnection
islands, which indicates that Fermi acceleration is more important
than direct acceleration at X-line. Once a particle travels into the
downstream region of the shock it rarely travels back into the upstream
region, indicating that Fermi mechanism in the reconnection islands
is also much more efficient than the diffusive shock acceleration
at least for the parameter regime we explore in this paper. The acceleration
pattern from particle trajectories is distinctly different from earlier
work by \citet{Sironi2011}, who concluded that direct acceleration
by non-ideal electric field at X-points is the main acceleration mechanism.

To distinguish the relative importance of Fermi acceleration and direct
acceleration, we present a number of analyses to statistically evaluate
the two mechanisms. The Fermi-type acceleration is mainly through
the electric field induced by bulk plasma motion, or more generally,
the electric field perpendicular to the local magnetic field, whereas
the direct acceleration is driven by the non-ideal electric field,
or parallel electric field if a non-zero magnetic field exists. One
can distinguish the two mechanisms by calculating the statistical
motion of the charged particles and the energization from different
components of electric fields. The generalized Fermi acceleration
in the relativistic reconnection layer \citep{Lemoine2019,Reconn_Guo2014,Reconn_Drake2006}
follows the particle journey through a sequence of local frames where
local electric field vanishes. For such local frames, we can decompose
the electric fields based on Lorentz transformation, with detailed
derivations given in the Appendix \ref{sec:Decomposition-f} and \ref{sec:Decomposition-g}.
The ways we decompose the electric fields are a generalization for
the non-relativistic case used extensively in previous studies \citep{Reconn_Guo2014,Reconn_Guo2015,Reconn_Guo2019,Reconn_Li2018}.
This generalization is critical in this work because the motion of
the bulk flow in our simulations can be highly relativistic. We also
calculate the integral of the energization from each term over $0<x<1500c/\omega_{p}$
and $0<y<400c/\omega_{p}$, and compare it with the integral of total
energization $P=\int\boldsymbol{J}\cdot\boldsymbol{E}$.

First, we distinguish the electric field associated with bulk plasma
motion (motional electric field) from the one that is not (non-ideal
electric field). As described in Appendix \ref{sec:Decomposition-f},
we use the following procedures: (1) make a Lorentz boost transforming
the electromagnetic field $\boldsymbol{E}$ and $\boldsymbol{B}$
in the simulation frame to $\boldsymbol{E}^{\prime}$ and $\boldsymbol{B}^{\prime}$
into the local co-moving frame (the frame moving at $\boldsymbol{\beta}_{f}c$,
where $\boldsymbol{\beta}_{f}c$ is the speed of the fluid motion),
(2) set $\boldsymbol{E}^{\prime}=0$ in the co-moving frame, (3) transform
the magnetic field $\boldsymbol{B}^{\prime}$ back into the simulation
frame as $\boldsymbol{E}_{m}$ and $\boldsymbol{B}_{m}$, where $\boldsymbol{E}_{m}$
is the electric field due to plasma motion and the remaining part
$\boldsymbol{E}_{n}=\boldsymbol{E}-\boldsymbol{E}_{m}$ is the non-ideal
electric field. In the non-relativistic limit, one obtains $\boldsymbol{E}_{m}\approx-\boldsymbol{\beta}_{f}\times\boldsymbol{B}$
and $\boldsymbol{E}_{n}\approx\boldsymbol{E}+\boldsymbol{\beta}_{f}\times\boldsymbol{B}$
\citep{Reconn_Guo2019}. The spatial profiles of $\boldsymbol{J}\cdot\boldsymbol{E}_{m}$
and $\boldsymbol{J}\cdot\boldsymbol{E}_{n}$ are shown in Figure \ref{fig:JdotE}(b)
and (c). The profiles of $\boldsymbol{J}\cdot\boldsymbol{E}$ and
$\boldsymbol{J}\cdot\boldsymbol{E}_{m}$ have similar features, both
showing different signs on different sides of each island, while $\boldsymbol{J}\cdot\boldsymbol{E}_{n}$
has smaller value but opposite polarity compared with $\boldsymbol{J}\cdot\boldsymbol{E}_{m}$.
We compare $P_{m}=\int\boldsymbol{J}\cdot\boldsymbol{E}_{m}$ with
$P=\int\boldsymbol{J}\cdot\boldsymbol{E}$ and find $P_{m}\approx0.56P$,
which indicates that the motional electric field has a larger contribution
to the particle energization than the non-ideal electric field.

The second way to decompose the electric field finds the generalized
perpendicular and parallel electric field through a series of frame
transformations. As described in Appendix \ref{sec:Decomposition-g},
we use the following procedures: (1) make a Lorentz boost transforming
the electromagnetic field $\boldsymbol{E}$ and $\boldsymbol{B}$
in the simulation frame to $\boldsymbol{E}^{\prime}$ and $\boldsymbol{B}^{\prime}$
in a local frame where $\boldsymbol{E}^{\prime}$ and $\boldsymbol{B}^{\prime}$
are parallel, generally this local frame is also the guiding center
frame (2) set $\boldsymbol{E}^{\prime}=0$ in that frame, (3) transform
the magnetic field $\boldsymbol{B}^{\prime}$ back into $\boldsymbol{E}_{g}$
and $\boldsymbol{B}_{g}$ in the simulation frame. Since $\boldsymbol{E}_{g}\cdot\boldsymbol{B}_{g}=0$
and $(\boldsymbol{E}-\boldsymbol{E}_{g})\parallel\boldsymbol{B}_{g}$,
we have the generalized perpendicular electric field $\boldsymbol{E}_{\perp}=\boldsymbol{E}_{g}$
and parallel electric field $\boldsymbol{E}_{\parallel}=\boldsymbol{E}-\boldsymbol{E}_{g}$.
In the non-relativistic limit we have $\boldsymbol{E}_{\perp}\approx\boldsymbol{E}-(\boldsymbol{E}\cdot\boldsymbol{B}/B^{2})\boldsymbol{B}$
and $\boldsymbol{E}_{\parallel}\approx(\boldsymbol{E}\cdot\boldsymbol{B}/B^{2})\boldsymbol{B}$
\citep{Reconn_Guo2014,Reconn_Guo2015,Reconn_Li2018}. The spatial
profiles of $\boldsymbol{J}\cdot\boldsymbol{E}_{\perp}$ and $\boldsymbol{J}\cdot\boldsymbol{E}_{\parallel}$
are shown in Figure \ref{fig:JdotE}(d) and (e). The spatial profile
and amplitude of $\boldsymbol{J}\cdot\boldsymbol{E}_{\perp}$ are
similar to that of $\boldsymbol{J}\cdot\boldsymbol{E}$ in Figure
\ref{fig:JdotE}(a), while $\boldsymbol{J}\cdot\boldsymbol{E}_{\parallel}$
is negligible compared with $\boldsymbol{J}\cdot\boldsymbol{E}_{\perp}$.
The integral over $0<x<1500c/\omega_{p}$ for $\boldsymbol{J}\cdot\boldsymbol{E}_{\perp}$
is $P_{\perp}=\int\boldsymbol{J}\cdot\boldsymbol{E}_{\perp}=0.84P$.
The fact that $P_{\parallel}=P-P_{\perp}=0.16P$ is small indicates
that particles gain energy mainly through Fermi-type acceleration.

We also compare the energization rate in $|\boldsymbol{E}|<|\boldsymbol{B}|$
region and $|\boldsymbol{E}|>|\boldsymbol{B}|$ region in Figure \ref{fig:JdotE}(f)
and (g) as suggested by \citet{Sironi2011,Reconn_Sironi2014}. Since
$|\boldsymbol{E}|^{2}-|\boldsymbol{B}|^{2}$ is a Lorentz invariant,
the decomposition into the $|\boldsymbol{E}|<|\boldsymbol{B}|$ region
and the $|\boldsymbol{E}|>|\boldsymbol{B}|$ region are consistent
among all the reference frames. Similar to the $\boldsymbol{J}\cdot(\boldsymbol{E}-\boldsymbol{E}_{g})$
term, the energization term in $|\boldsymbol{E}|>|\boldsymbol{B}|$
region is negligible. By integrating $\boldsymbol{J}\cdot\boldsymbol{E}$
over subregions where $|\boldsymbol{E}|<|\boldsymbol{B}|$, we find
that $P_{L}=\int_{|\boldsymbol{E}|<|\boldsymbol{B}|}\boldsymbol{J}\cdot\boldsymbol{E}=0.93P$.
The fact that $P-P_{L}=0.07P$ is small confirms that the energy conversion
from the X-points is much weaker than that in the reconnection islands
\citep{Reconn_Guo2015,Reconn_Guo2019}.

We further study the averaged acceleration rate $\langle W\rangle\equiv\langle q\boldsymbol{v}\cdot\boldsymbol{E}\rangle$
for particles of different energy in the simulation at $\omega_{p}t=2000$.
We show $\langle W\rangle$ in three different regions, in Figure
\ref{fig:plot-acc-a}(a) and Figure \ref{fig:plot-acc-b}(a) for region
I ($0<x<197c/\omega_{p}$, where the reconnection islands are in the
downstream and already coalesce and grow to large size), in Figure
\ref{fig:plot-acc-a}(b) and Figure \ref{fig:plot-acc-b}(b) for region
II ($573c/\omega_{p}<x<804c/\omega_{p}$, where the magnetic islands
have a large size but have not yet reached the main shock), and in
Figure \ref{fig:plot-acc-a}(c) and Figure \ref{fig:plot-acc-b}(c)
for region III ($1275c/\omega_{p}<x<1497c/\omega_{p}$, where a current
sheet just breaks into magnetic islands). Region I is in the downstream
of the shock, while region II and III are a pre-shock regions with
on-going magnetic reconnection. These regions are also marked as in
Figure \ref{fig:plot2D}(d) and Figure \ref{fig:JdotE}. The averaged
acceleration rate is roughly proportional to particle energy over
a wide range of energies, indicating that the major acceleration mechanism
is Fermi acceleration. The non-ideal electric field $\boldsymbol{E}_{n}$
only contribute to positive acceleration in a small subset of energy
ranges especially in low energies compared to the contribution from
the motional electric field $\boldsymbol{E}_{m}$. The parallel electric
field $\boldsymbol{E}_{\parallel}$ only contribute to positive acceleration
in a small subset of energy ranges in high energies compared to the
contribution from the motional electric field $\boldsymbol{E}_{\perp}$.

Theoretical analysis \citep{Reconn_Guo2014,Reconn_Guo2015,Reconn_Guo2019}
has shown that the power-law of particle spectrum can be explained
by solving the energy continuity equation with injection of particles.
While Fermi acceleration does not change the shape of the spectrum
for the particles in the initial reconnection layer, the injected
particles into the layer can form a power law distribution with accelerated
energy. In the case of the relativistic striped wind, the injection
is continuously going on due to the incoming flow from the striped
wind and the particles flowing into the downstream of the shock. The
first-order Fermi process, where the acceleration rate is proportional
to the energy of the particles as confirmed in Figure \ref{fig:plot-acc-a},
is accompanied by the escape of particles from the reconnection islands.
In the case of $\lambda=640d_{e}$, the spectrum we get from the simulation
has $p=1.5$ as shown in Figure \ref{fig:plot-spect1}, which is consistent
with $1<p<2$ as we get from the analytical model \citep{Reconn_Guo2014,Reconn_Guo2015,Reconn_Guo2019}.
In the case of smaller wavelength $\lambda\lesssim40d_{e}$ for the
striped wind, there are more X-points as the shock-reconnection system
evolves. However, the magnetic islands occupy a larger fraction of
the total area for a smaller wavelength for the striped wind, resulting
in more particles in the initial reconnection layer and less particles
injected into the layer. Thus, by applying the theoretical analysis
in \citet{Reconn_Guo2014} the particle spectrum are softer and more
Maxwellian-like as the wavelength $\lambda$ decreases, as shown in
Figure \ref{fig:plot-spect2}(b).

\section{Conclusions and discussions}

\label{sec:Conclusions-and-discussions}

By carrying out first principles kinetic PIC simulations, we have
studied the shock structure and dynamics, magnetic reconnection, and
particle energization at the termination shock of a relativistic striped
wind. Our parameter regime covers an unprecedented relativistic regime
with large bulk Lorentz factor up to $\gamma_{0}=10^{6}$, which are
expected in PWNe. The values of $\gamma_{0}$ and $\sigma_{0}$ are
even uncertain for a single PWN, due to the uncertainties in the density
at the light cylinder and the global dissipation mechanism in the
nebula \citep{Striped_Kirk2003,Flares_Kirk2017}. For the observed
gamma-ray flares \citep{Flares_Tavani2011,Flares_Abdo2011,Flares_Buehler2012a,Flares_Weisskopf2013,Flares_Buehler2014}
in the Crab nebula, a sudden drop in the mass loading of the pulsar
wind may result in $\sigma_{0}\approx10$ and $\gamma_{0}>10^{6}$
at the termination shock \citep{Flares_Kirk2017}. The maximum energy
for accelerated electrons and positrons can reach hundreds of TeV
if $\gamma_{0}\approx10^{6}$ and $\sigma_{0}=10$. Our study is relevant
to magnetic energy conversion and particle acceleration in PWNs produced
by oblique rotating pulsars. While a growing body of research using
PIC methods focuses on particle acceleration in the spontaneous relativistic
magnetic reconnection in the magnetically dominated regime, in this
work we extended the study by examining the particle acceleration
mechanism in shock driven magnetic reconnection at the termination
shock of relativistic striped wind. While the plasma dynamics in this
regime is studied extensively by \citet{Sironi2011}, we study carefully
the particle energization and acceleration. By analyzing PIC simulations,
we found that the termination shock is efficient at converting magnetic
energy to particle energy and producing non-thermal power law distributions.
The particle distribution is scalable with upstream bulk Lorentz factor
$\gamma_{0}$ and magnetization $\sigma_{0}$. For sufficiently large
$\gamma_{0}$ and $\sigma_{0}$, the spectrum we get from the simulation
has a power law with $p=1.5$ in energy range $0.2<\gamma/(\gamma_{0}\sigma_{0})<4$
for the standard run. Detailed analysis shows that Fermi-type mechanism
dominates the particle acceleration and power-law formation. For smaller
wavelength ($\lambda\lesssim40d_{e}$ for $\sigma_{0}=10$), the spectrum
is more Maxwellian-like due to more particles in the initial reconnection
layer and less particles injected into the layer. 

The high energy particles are known to efficiently produce radiation
through a few mechanisms, e.g. synchrotron, bremsstrahlung and inverse-Compton
scattering. The morphology of magnetic fields and the spectra of electron-positron
pair plasma we get from the PIC simulations can be post-processed
to generate the spectrum \citep{Lyutikov2019} and polarization \citep{POL_Dean2008,POL_Forot2008,POL_Jourdain2019}
of the resulting radiation and compared to the observations from the
Crab and other PWNe, which will be subject of future reports.

\section*{Acknowledgements}

Research presented in this paper was supported by the Center for Space
and Earth Science (CSES) program and Laboratory Directed Research
and Development (LDRD) program 20200367ER of Los Alamos National Laboratory
(LANL). The research by PK was also supported by CSES program. The
research by YL was supported by CSES student fellowship project. CSES
is funded by LANL's LDRD program under project number 20180475DR.
FG, PK and HL also acknowledge support from DOE through the LDRD program
at LANL and DOE/OFES support to LANL, and NASA Astrophysics Theory
Program. The simulations were performed with LANL Institutional Computing
which is supported by the U.S. Department of Energy National Nuclear
Security Administration under Contract No. 89233218CNA000001, and
with the Extreme Science and Engineering Discovery Environment (XSEDE),
which is supported by National Science Foundation (NSF) grant number
ACI-1548562.

\bibliographystyle{apsrev}
\bibliography{PWNE}

\appendix

\section{Test problem with double-periodic box}

\label{sec:Test-problem-periodic}

\begin{figure*}[tph]
\includegraphics[scale=0.7]{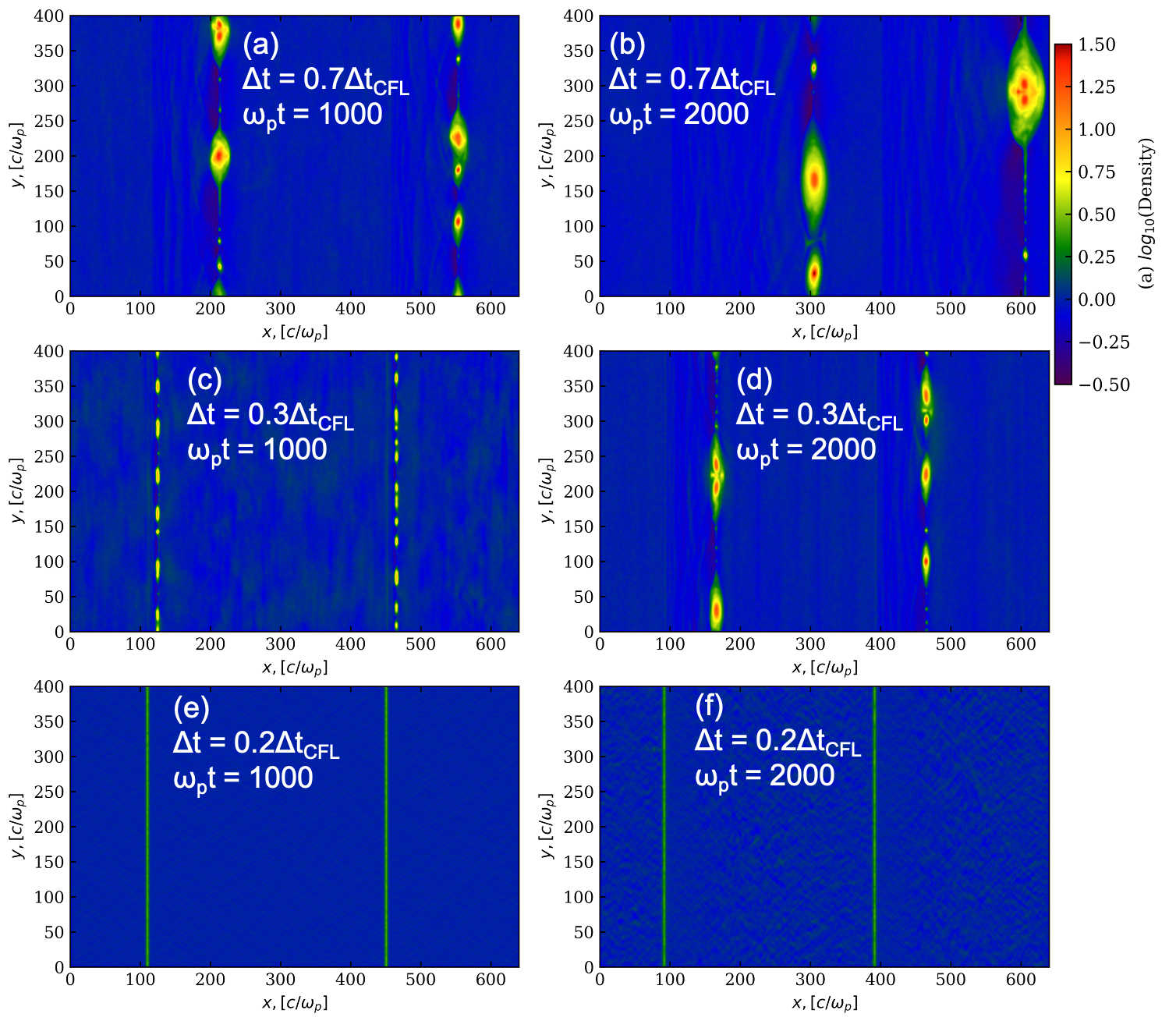}

\caption{The results for the test problem with $\Delta=d_{e}$, $L_{x}=\lambda=640d_{e}$
and $L_{y}=400d_{e}$ using double-periodic box. The variable plotted
is $\log_{10}(n/n_{c0})$. Subfigures (a), (c) and (e) are at $\omega_{p}t=1000$,
and subfigures (b), (d) and (f) are at $\omega_{p}t=2000$. For different
time steps, $\Delta t=0.7\Delta t_{\mathrm{CFL}}$ in (a) and (b),
$\Delta t=0.3\Delta t_{\mathrm{CFL}}$ in (c) and (d), and $\Delta t=0.2\Delta t_{\mathrm{CFL}}$
in (e) and (f). The maximum stable time step under the CFL condition
is given by $\Delta t_{\mathrm{CFL}}=1/(c\sqrt{1/\Delta x^{2}+1/\Delta y^{2}})$.
\label{fig:Results-for-test}}
\end{figure*}

To make sure the relativistic reconnection onset in our simulations
is not due to numerical effects, we carry out several test runs for
a current sheet moving in a double-periodic box. The simulation setup
is similar to run S0, with $\sigma_{0}=10$, $\alpha=0.1$, $L_{x}=\lambda=640d_{e}$,
$L_{y}=400d_{e}$ and $\Delta=d_{e}$, except that the $x$ direction
is periodic and only covers one wavelength of the striped wind. Therefore,
the simulation tests the stability of the striped wind. We use the
force interpolation scheme called the WT (standing for \textbf{w}eighting
with \textbf{t}ime-step dependency) scheme proposed in a recent development
\citep{NCI_Lu2020} to overcome the numerical Cherenkov instability
(NCI). The WT scheme requires a small time step to improve the numerical
stability of the simulations. We test three cases with different time
steps $\Delta t=0.7\Delta t_{\mathrm{CFL}}$, $\Delta t=0.3\Delta t_{\mathrm{CFL}}$
and $\Delta t=0.2\Delta t_{\mathrm{CFL}}$, where $\Delta t_{\mathrm{CFL}}=1/(c\sqrt{1/\Delta x^{2}+1/\Delta y^{2}})$.

The results for the test problem are shown in Figure \ref{fig:Results-for-test}.
According to the growth rate of tearing instability given in a previous
study \citep{Reconn_Galeev1986,Reconn_Daughton2011}, the current
sheets in the test runs should not reconnect when $\omega_{p}t$ is
a few thousand. Any current sheets reconnect before that time is caused
by numerical problems. For the runs with $\Delta t=0.7\Delta t_{\mathrm{CFL}}$
and $\Delta t=0.3\Delta t_{\mathrm{CFL}}$, the reconnection happens
before $\omega_{p}t<1000$ and the islands have significantly grown
their size at $\omega_{p}t=2000$, which is roughly the time it takes
for the rightmost current sheet to travel before it reaches the main
shock. Thus, the results for the runs with $\Delta t=0.7\Delta t_{\mathrm{CFL}}$
and $\Delta t=0.3\Delta t_{\mathrm{CFL}}$ are not consistent with
the theoretical prediction. The theoretical prediction \citep{Reconn_Galeev1986,Reconn_Daughton2011}
is verified using the runs in the co-moving frame of the flow. For
the run with $\Delta t=0.2\Delta t_{\mathrm{CFL}}$, the current sheets
have not significantly change their morphology from the initial condition
until $\omega_{p}t=2000$. Since the run with $\Delta t=0.2\Delta t_{\mathrm{CFL}}$
is the most stable run in Figure \ref{fig:Results-for-test}, and
it can make sure the current sheet is stable before it interacts with
the fast shock, we use the small time step $\Delta t=0.2\Delta t_{\mathrm{CFL}}$
in all the production runs shown in the main context.

\section{Scaling equations for relativistic magnetized plasmas}

\label{sec:Scaling-equations-for}

We show that the particle-field equations for relativistic PIC modeling
can be written in the dimensionless form with proper normalization.
The equations for relativistic PIC modeling in cgs units are

\begin{equation}
\begin{aligned}m_{s}\frac{d\boldsymbol{u}_{s}}{dt} & =q_{s}(\boldsymbol{E}+\frac{\boldsymbol{v}_{s}}{c}\times\boldsymbol{B})\\
\frac{d\boldsymbol{x}_{s}}{dt} & =\boldsymbol{v}_{s}\\
\frac{\partial\boldsymbol{E}}{\partial t} & =c\nabla\times\boldsymbol{B}-4\pi\boldsymbol{J}\\
\frac{\partial\boldsymbol{B}}{\partial t} & =-c\nabla\times\boldsymbol{E}\\
\boldsymbol{J} & =\sum_{s}w_{s}q_{s}\boldsymbol{v}_{s}
\end{aligned}
\end{equation}
where $s$ stands for $s$-th quasi-particle and $w_{s}$ is the weight
of $s$-th pseudo-particle. We introduce the following normalization
\begin{equation}
\begin{array}{cc}
 & t=\sigma_{0}^{1/2}\omega_{pe}^{-1}\tilde{t},\qquad\boldsymbol{x}_{s}=\sigma_{0}^{1/2}(c/\omega_{p})\tilde{\boldsymbol{x}},\\
 & \boldsymbol{u}_{s}=\sigma_{0}\gamma_{0}c\tilde{\boldsymbol{u}}_{s},\qquad\boldsymbol{v}_{s}=c\tilde{\boldsymbol{v}}_{s},\\
 & m_{s}=m_{e}\tilde{m}_{s},\qquad q_{s}=e\tilde{q}_{s},\qquad\tilde{w}_{s}=n_{c0}\tilde{n},\\
 & \boldsymbol{E}=\sqrt{4\pi\gamma_{0}n_{c0}m_{e}c^{2}\sigma_{0}}\tilde{\boldsymbol{E}}\qquad\boldsymbol{B}=\sqrt{4\pi\gamma_{0}n_{c0}m_{e}c^{2}\sigma_{0}}\tilde{\boldsymbol{B}}
\end{array}\label{eq:norm}
\end{equation}
where $\omega_{p}=\sqrt{4\pi n_{c0}e^{2}/(\gamma_{0}m_{e})}$, then
we obtain the dimensionless equations
\begin{equation}
\begin{aligned}\frac{d\tilde{\boldsymbol{u}}_{s}}{d\tilde{t}} & =\frac{\tilde{q}_{s}}{\tilde{m}_{s}}(\tilde{\boldsymbol{E}}+\tilde{\boldsymbol{v}}_{s}\times\tilde{\boldsymbol{B}})\\
\frac{d\tilde{\boldsymbol{x}}_{s}}{d\tilde{t}} & =\frac{\tilde{\boldsymbol{v}}_{s}}{c}\\
\frac{\partial\tilde{\boldsymbol{E}}}{\partial\tilde{t}} & =\tilde{\nabla}\times\tilde{\boldsymbol{B}}-\sum_{s}\tilde{w}_{s}\tilde{q}_{s}\tilde{\boldsymbol{v}}_{s}\\
\frac{\partial\tilde{\boldsymbol{B}}}{\partial\tilde{t}} & =-\tilde{\nabla}\times\tilde{\boldsymbol{E}}
\end{aligned}
\label{eq:scale-eq1}
\end{equation}
and 
\begin{equation}
\tilde{\boldsymbol{v}}=\frac{1}{\sqrt{1+1/(\tilde{u}^{2}\gamma_{0}^{2}\sigma_{0}^{2})}}\frac{\tilde{\boldsymbol{u}}}{\tilde{u}}=(1-\frac{1}{2\tilde{u}^{2}\gamma_{0}^{2}\sigma_{0}^{2}}+\mathcal{O}(\frac{1}{\tilde{u}^{4}\gamma_{0}^{4}\sigma_{0}^{4}}))\frac{\tilde{\boldsymbol{u}}}{\tilde{u}}\label{eq:v-and-u}
\end{equation}
As long as $\gamma_{0}\sigma_{0}\tilde{u}\gg1$, Eq(\ref{eq:v-and-u})
can be well approximated by $\tilde{\boldsymbol{v}}=\tilde{\boldsymbol{u}}/\tilde{u}$,
then the dimensionless equations Eq(\ref{eq:scale-eq1}) are the same
for different $\gamma_{0}$ and $\sigma_{0}$. For the simulations
in this work, in the upstream flow, we have $\tilde{u}=u/(\sigma_{0}\gamma_{0}c)\approx1/\sigma_{0}\ll1$,
$\tilde{E}\approx\tilde{B}=1$, and the normalized wavelength $\tilde{\lambda}=\lambda/(\sqrt{\sigma_{0}}c/\omega_{p})$.
As long as the $\tilde{\lambda}$ keeps the same, no matter what $\sigma_{0}$
and $\gamma_{0}$ we have, the initial conditions and the solution
for Eq(\ref{eq:scale-eq1}) are scalable. One can then use the solution
of Eq(\ref{eq:scale-eq1}) to obtain scalable dimensional results
using normalization factors from Eq(\ref{eq:norm}) .

\section{Decomposing electromagnetic fields}

The generalized Fermi acceleration \citep{Lemoine2019} follows the
particle journey through a sequence of local frames where local electric
field vanishes. To find such local frames, we can decompose the electric
fields based on Lorentz transformation. We generalize the decomposition
in previous studies \citep{Reconn_Guo2014,Reconn_Guo2015,Reconn_Guo2019,Reconn_Li2018}
to the relativistic case, since the bulk flow in this work can be
highly relativistic. In Appendix \ref{sec:Decomposition-f}, by removing
the electric field in the fluid co-moving frame, we distinguish the
electric field associated with bulk plasma motion (motional electric
field) from the one that is not (non-ideal electric field). In Appendix
\ref{sec:Decomposition-g}, by removing the electric field in the
frame where the electric and magnetic fields are parallel, we get
the generalized parallel and perpendicular electric fields.

\subsection{Decomposing electromagnetic field in the bulk fluid frame}

\label{sec:Decomposition-f}

For convenience, we define the complex vector of electromagnetic field
as $\boldsymbol{F}=\boldsymbol{E}+i\boldsymbol{B}$ \citep{Landau1963}.
The Lorentz transformation for $\boldsymbol{F}=\boldsymbol{E}+i\boldsymbol{B}$
in the simulation frame $S$ to $\boldsymbol{F}^{\prime}=\boldsymbol{E}^{\prime}+i\boldsymbol{B}^{\prime}$
in fluid co-moving frame $S_{f}$ which moves at speed $\boldsymbol{\beta}_{f}c$
with respect to $S$ is (see Chapter 26 in \citealp{Feynman1964})
\begin{equation}
\boldsymbol{F}^{\prime}=(1-\gamma_{f})(\boldsymbol{F}\cdot\boldsymbol{n}_{f})\boldsymbol{n}_{f}+\gamma_{f}(\boldsymbol{F}-i\boldsymbol{\beta}_{f}\times\boldsymbol{F})
\end{equation}
where $\boldsymbol{n}_{f}=\boldsymbol{\beta}_{f}/\beta_{f}$ is the
direction of $\boldsymbol{\beta}_{f}$, and $\gamma_{f}=1/\sqrt{1-\beta_{f}^{2}}$.
If we have a field configuration $\boldsymbol{F}_{m}^{\prime}$ in
$S_{f}$ frame such that the electric field $\boldsymbol{E}_{m}^{\prime}$
is zero and the magnetic field $\boldsymbol{B}_{m}^{\prime}$ is same
as $\boldsymbol{B}^{\prime}$, then
\begin{align}
\boldsymbol{F}_{m}^{\prime} & =i\boldsymbol{B}^{\prime}\nonumber \\
 & =i[(1-\gamma_{f})(\boldsymbol{B}\cdot\boldsymbol{n}_{f})\boldsymbol{n}_{f}+\gamma_{f}(\boldsymbol{B}-\boldsymbol{\beta}_{f}\times\boldsymbol{E})]
\end{align}
Transform back $\boldsymbol{F}_{m}^{\prime}$ into $\boldsymbol{F}_{m}$
in $S$ frame, we obtain
\begin{align}
\boldsymbol{F}_{m} & =(1-\gamma_{f})(\boldsymbol{F}_{m}^{\prime}\cdot\boldsymbol{n}_{f})\boldsymbol{n}_{f}+\gamma_{f}(\boldsymbol{F}_{m}^{\prime}+i\boldsymbol{\beta}_{f}\times\boldsymbol{F}_{m}^{\prime})\nonumber \\
 & =[-\gamma_{f}^{2}\boldsymbol{\beta}_{f}\times\boldsymbol{B}-(\gamma_{f}^{2}-1)(\boldsymbol{E}-(\boldsymbol{n}_{f}\cdot\boldsymbol{E})\boldsymbol{n}_{f})]\nonumber \\
 & \qquad+i[-\gamma_{f}^{2}\boldsymbol{\beta}_{f}\times\boldsymbol{E}\nonumber \\
 & \qquad+\gamma_{f}^{2}(\boldsymbol{B}-(\boldsymbol{B}\cdot\boldsymbol{n}_{f})\boldsymbol{n}_{f})+(\boldsymbol{B}\cdot\boldsymbol{n}_{f})\boldsymbol{n}_{f}]
\end{align}
The real part of $\boldsymbol{F}_{m}$, i.e. the electric field due
to plasma motion, is 
\begin{align}
\boldsymbol{E}_{m} & =-\gamma_{f}^{2}(\boldsymbol{\beta}_{f}\times\boldsymbol{B}+\boldsymbol{E}-(\boldsymbol{n}_{f}\cdot\boldsymbol{E})\boldsymbol{n}_{f})+(\boldsymbol{E}-(\boldsymbol{n}_{f}\cdot\boldsymbol{E})\boldsymbol{n}_{f})\nonumber \\
 & =\frac{(\boldsymbol{\mathcal{P}}\cdot\boldsymbol{E})\boldsymbol{\mathcal{\mathcal{P}}}-\mathcal{\mathcal{P}}^{2}\boldsymbol{E}-(\mathcal{E}/c)\boldsymbol{\mathcal{P}}\times\boldsymbol{B}}{(\mathcal{E}/c)^{2}-\mathcal{P}^{2}}\nonumber \\
 & =\frac{\boldsymbol{\mathcal{\mathcal{P}}}\times[\boldsymbol{\mathcal{\mathcal{P}}}\times\boldsymbol{E}-(\mathcal{E}/c)\boldsymbol{B}]}{(\mathcal{E}/c)^{2}-\mathcal{P}^{2}}
\end{align}
where $\boldsymbol{\mathcal{\mathcal{P}}}$ is the momentum density
of the fluid and $\mathcal{E}$ is the energy density of the fluid,
and the fluid velocity is given by $\boldsymbol{\beta}_{f}c=\boldsymbol{\mathcal{\mathcal{P}}}c^{2}/\mathcal{E}$.
In the case where $\beta_{f}\ll1$, we have $\boldsymbol{E}_{m}\approx-\boldsymbol{\beta}_{f}\times\boldsymbol{B}$.

\subsection{Decomposing electromagnetic field in a frame where electric and magnetic
fields are parallel}

\label{sec:Decomposition-g}

A frame where the electric and magnetic fields are parallel can be
generally found for given electromagnetic field $(\boldsymbol{E},\boldsymbol{B})$
as long as $E^{2}+B^{2}>0$, even for the $B=0$ case. The velocity
$\boldsymbol{v}_{g}=\boldsymbol{\beta}_{g}c$ of such frame is given
by (see \citealp{EM_Petri2019} or Section 25 in \citealp{Landau1963})

\begin{equation}
\frac{\boldsymbol{\beta}_{g}}{1+\beta_{g}^{2}}=\frac{\boldsymbol{E}\times\boldsymbol{B}}{E^{2}+B^{2}}\label{eq:beta-g}
\end{equation}
where $\boldsymbol{\beta}_{g}$ and the corresponding Lorentz factor
$\gamma_{g}$ can be solved as 
\begin{align}
\boldsymbol{\beta}_{g} & =\frac{2\boldsymbol{E}\times\boldsymbol{B}}{E^{2}+B^{2}+\sqrt{(E^{2}-B^{2})^{2}+4|\boldsymbol{E}\cdot\boldsymbol{B}|^{2}}}\\
\gamma_{g}^{2} & =\frac{E^{2}+B^{2}+\sqrt{(E^{2}-B^{2})^{2}+4|\boldsymbol{E}\cdot\boldsymbol{B}|^{2}}}{2\sqrt{(E^{2}-B^{2})^{2}+4|\boldsymbol{E}\cdot\boldsymbol{B}|^{2}}}
\end{align}

The electromagnetic field $(\boldsymbol{E},\boldsymbol{B})$ in the
simulation frame is transformed into the electromagnetic field $(\boldsymbol{E}^{\prime},\boldsymbol{B}^{\prime})$
in frame $S_{g}$ through a Lorentz transformation using $\boldsymbol{v}_{g}$.
In frame $S_{g}$, the electromagnetic field $(\boldsymbol{E}^{\prime},\boldsymbol{B}^{\prime})$
satisfies $\boldsymbol{E}^{\prime}\parallel\boldsymbol{B}^{\prime}$
and the particle motion follows an accelerating motion in $\boldsymbol{B}^{\prime}$
direction and a gyro-motion perpendicular to $\boldsymbol{B}^{\prime}$
direction.

The complex scalar 
\begin{equation}
\boldsymbol{F}^{2}=\boldsymbol{F}\cdot\boldsymbol{F}=(E^{2}-B^{2})+2i\boldsymbol{E}\cdot\boldsymbol{B}=C_{R}+iC_{I}
\end{equation}
is a Lorentz invariant, where $C_{R}=E^{2}-B^{2}$ and $C_{I}=2\boldsymbol{E}\cdot\boldsymbol{B}$
\citep{Landau1963}. The Lorentz transform for $\boldsymbol{F}$ in
the simulation frame $S$ to $\boldsymbol{F}^{\prime}=\boldsymbol{E}^{\prime}+i\boldsymbol{B}^{\prime}$
in a frame $S^{\prime}$ that moves at speed $\boldsymbol{\beta}_{g}c$
with respect to $S$ is
\begin{equation}
\boldsymbol{F}^{\prime}=(1-\gamma_{g})(\boldsymbol{F}\cdot\boldsymbol{n}_{g})\boldsymbol{n}_{g}+\gamma_{g}(\boldsymbol{F}-i\boldsymbol{\beta}_{g}\times\boldsymbol{F})
\end{equation}
where $\boldsymbol{n}_{g}=\boldsymbol{\beta}_{g}/\beta_{g}$ is the
direction of $\boldsymbol{\beta}_{g}$. If we define a field configuration
$\boldsymbol{F}_{g}^{\prime}$ in $S^{\prime}$ frame such that the
electric field $\boldsymbol{E}_{g}^{\prime}$ is zero and the magnetic
field $\boldsymbol{B}_{g}^{\prime}$ is same as $\boldsymbol{B}^{\prime}$,
then by transforming back $\boldsymbol{F}_{g}^{\prime}$ into $\boldsymbol{F}_{g}$
in $S$ frame, we obtain

\begin{align}
\boldsymbol{F}_{g} & =\gamma_{g}[\boldsymbol{F}_{a}^{\prime}+i\boldsymbol{\beta}_{g}\times\boldsymbol{F}_{a}^{\prime}]\nonumber \\
 & =-\gamma_{g}^{2}\boldsymbol{\beta}\times(\boldsymbol{B}-\boldsymbol{\beta}_{g}\times\boldsymbol{E})+i\gamma_{g}^{2}(\boldsymbol{B}-\boldsymbol{\beta}_{g}\times\boldsymbol{E})\nonumber \\
 & =(\frac{1}{2}-\frac{C_{R}}{2\sqrt{C_{R}^{2}+C_{I}^{2}}}+\frac{C_{I}i}{2\sqrt{C_{R}^{2}+C_{I}^{2}}})(\boldsymbol{E}+i\boldsymbol{B})
\end{align}
or
\begin{equation}
\begin{aligned}\boldsymbol{E}_{g} & =(\frac{1}{2}-\frac{C_{R}}{2\sqrt{C_{R}^{2}+C_{I}^{2}}})\boldsymbol{E}-(\frac{C_{I}}{2\sqrt{C_{R}^{2}+C_{I}^{2}}})\boldsymbol{B}\\
\boldsymbol{B}_{g} & =(\frac{1}{2}-\frac{C_{R}}{2\sqrt{C_{R}^{2}+C_{I}^{2}}})\boldsymbol{B}+(\frac{C_{I}}{2\sqrt{C_{R}^{2}+C_{I}^{2}}})\boldsymbol{E}
\end{aligned}
\end{equation}
Since $\boldsymbol{E}_{g}\cdot\boldsymbol{B}_{g}=0$ and $(\boldsymbol{E}-\boldsymbol{E}_{g})\parallel\boldsymbol{B}_{g}$,
we have the generalized perpendicular electric field $\boldsymbol{E}_{\perp}=\boldsymbol{E}_{g}$
and generalized parallel electric field $\boldsymbol{E}_{\parallel}=\boldsymbol{E}-\boldsymbol{E}_{g}$.
In the case where $|\boldsymbol{E}|\ll|\boldsymbol{B}|$ we have $|C_{I}/C_{R}|\approx2\boldsymbol{E}\cdot\boldsymbol{B}/B^{2}\ll1$
and $C_{R}<0$, thus the transformation is equivalent to removing
the parallel electric field and keeping the perpendicular electric
field
\begin{equation}
\begin{aligned}\boldsymbol{E}_{\mathrm{\perp}}\approx & \boldsymbol{E}-\frac{\boldsymbol{E}\cdot\boldsymbol{B}}{B^{2}}\boldsymbol{B}\\
\boldsymbol{E}_{\mathrm{\parallel}}\approx & \frac{\boldsymbol{E}\cdot\boldsymbol{B}}{B^{2}}\boldsymbol{B}\\
\boldsymbol{B}_{g}\approx & \boldsymbol{B}
\end{aligned}
\end{equation}

\end{document}